\documentclass[global]{svjour}

\usepackage[latin1]{inputenc}
\usepackage[T1]{fontenc}
\usepackage{epsfig}
\usepackage{graphicx}
\usepackage{epstopdf}
\usepackage[authoryear,square]{natbib}
\usepackage{multirow}

\newcommand{\ie}{{\em i.e., }}           % i.e.
\newcommand{\eg}{{\em e.g., }}           % e.g.
\renewcommand{\d}{{\mathrm d}}             % differential

\def\be{\begin{equation}}
\def\ee{\end{equation}}
\def\ba{\begin{array}}
\def\ea{\end{array}}

\begin{document}
\title{Material-independent crack arrest statistics:
Application to indentation experiments}
\author{Yann Charles\inst{1,2} \and
Fran\c{c}ois Hild\inst{1}\thanks{to whom correspondence should be
addressed. Email: hild@lmt.ens-cachan.fr, Fax:~+33~1~47~40~22~40.}
\and St\'ephane Roux\inst{3} \and Damien Vandembroucq\inst{3}\\}

\institute{\inst{1}
LMT-Cachan, ENS de Cachan / CNRS-UMR 8535 / Universit\'{e} Paris 6\\
61 avenue du Pr\'{e}sident Wilson, F-94235 Cachan Cedex, France.\\
\\
\inst{2}
Now at GEMPPM, INSA-Lyon / CNRS-UMR 5510 \\
20 avenue Albert Einstein, F-69621 Villeurbannes Cedex, France.\\
\\
\inst{3} Unit\'{e} Mixte CNRS/Saint-Gobain, Surface du Verre et Interfaces \\
39 quai Lucien Lefranc, F-93303 Aubervilliers Cedex, France.}

\titlerunning{Crack arrest statistics}
\maketitle

\begin{abstract}
An extensive experimental study of indentation and crack arrest
statistics is presented for four different brittle materials
(alumina, silicon carbide, silicon nitride, glass). Evidence is
given that the crack length statistics is described by a
universal (\ie material independent) distribution. The latter
directly derives from results obtained when modeling crack
propagation as a depinning phenomenon. Crack arrest (or effective
toughness) statistics appears to be fully characterized by two
parameters, namely, an asymptotic crack length (or macroscopic
toughness) value and a power law size dependent width. The
experimental knowledge of the crack arrest statistics at one given
scale thus gives access to its knowledge at all scales.
\end{abstract}
\keywords{Brittle fracture, Crack arrest, Heterogeneity, Depinning,
Indentation, Probability and statistics.}

\newpage

\section{Introduction}

\begin{figure}[ht!]
\begin{center}
\epsfig{file=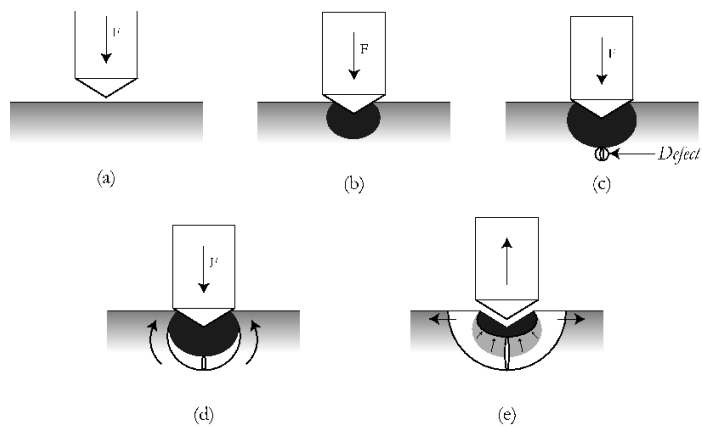,width=0.8\hsize}
\end{center}
\vfill \caption{Indentation principle. At the beginning of the
indentation test, a plastic zone is created below the Vickers
pyramid (b), inducing residual stresses. When the total stress
(\ie the applied one and the residual one) is large enough, two
perpendicular elementary cracks are created at the deepest
location under the plastic zone (c). For brittle media, it is
admitted that cracks are initiated from material defects. These
two cracks propagate along the plastic zone (d) when the load is
about to reach its maximum value. Then, while unloading the
sample, cracks finish their propagation, and their final form
is semi-circular (e). } \label{fig:principe}
\end{figure}

Indentation tests are widely used to estimate material properties
such as hardness and toughness. These tests are local and non
destructive. The application on the surface of a diamond pyramid
with a force $F$ creates an irreversible mark for any type of
material and a network of cracks for brittle solids~\citep{1431}
whose shape and direction depend on the tip geometry (Knoop,
Vickers, Berkovich \--- see Fig.~\ref{fig:principe} for a Vickers
indentor). The projected area $A$ of the permanent print defines
the Meyer hardness $H=F/A$. The toughness measurement mainly
relies on the existence on an intermediate scaling law between
the stress intensity factor (SIF) $K$ and the length $c$ of the
cracks generated during the test $K\propto F/c^{3/2}$ with a
geometry and material dependent prefactor. This scaling is easily
recovered in the framework of linear elastic fracture mechanics.
Assuming a point force $F$ (thus a stress field $F/r^2$), the
elastic energy released in a volume $c^3$ due to the presence of
radial cracks of radius $c$ is estimated to be $F^2/Ec$. The
energy release rate associated with a surface increase $\d
S\propto c\d c$ is thus $G=K^2/E\propto F^2/Ec^3$ so that
$K\propto F/c^{3/2}$.

The propagation being stable, the indentation cracks
are arrested for a length such that the SIF equals the toughness
$K=K_c$. Simple dimensional analysis shows that the above scaling
holds when the plastic strain is neglected. However, even when
plasticity is considered a similar scaling may hold.
\citet{Lawn-JACS80} used a plastic cavity model to estimate the
residual stresses induced by indentation and proposed the
following estimate for the toughness
$$
K_c=\chi_R\frac{F}{c^{3/2}} \;,\quad \chi_R=\xi_0 (\cot \Phi
)^{2/3} \sqrt{\frac{E}{H}}
$$
where $E$ and $H$ are respectively the Young's modulus and the
hardness, and $\Phi$ is the half-angle of the indentor. Last,
$\xi_0$ is a dimensionless constant that remains slightly
dependent on the plastic constitutive law of the material. Ample
experimental data show that for large loads, such a scaling
relation between $F$ and $c$ holds~\citep{1241}. In the case of a
Vickers indentation, one usually considers that this is the case
for radial cracks for which $c/a>2$, where $a$ is the half diagonal
of the plastic mark and $c$ is the crack length.
For smaller cracks, %(\ie smaller loads),
(\ie for smaller $c/a$ ratio) deviations are systematically
observed, but among the long list of proposed
relations~\citep{1241} between $F$ and $c$ no general conclusions
are derived in this regime. Note that for such lengths,
possible Palmqvist (or lateral) cracks may have been generated.

A key question has been raised in particular
by~\citet{Cook-JACS85}. Is it valid to extrapolate the toughness
value obtained at a microscopic scale with the indentation test to
the large scale of macroscopic cracks?  One limitation that is
easily be foreseen is the effect of microstructure. At a small
scale, the heterogeneous nature of the material may give rise to a
local toughness variability (\eg nature of the phases, orientation
of grains, grain boundaries, intrinsic heterogeneity).  What will
be the macroscopic consequences of such a variability?  A
phenomenological approach to this question consists in introducing
the so-called $R$-curve effect, where the apparent toughness
depends explicitly on the crack length. \citet{Cook-JACS85}
proposed to introduce a shielding effect at the scale of the
microstructure to give a phenomenological account to this
phenomenon.

Beyond this systematic size effect, the local disorder induces a
statistical scatter of the toughness measurements. Such a statistical
scatter is well known in the context of strength
measurements~\citep{47} and appears at the initiation step.  The
fluctuations thus reflect the defect distribution within the
material~\citep{80}, especially at its surface for
glass~\citep{1354}. In the case of indentation with a sharp tip,
initiation usually takes place below the surface on defects which are
either structural or induced by the plastic deformation.

The propagation being stable \ie the
SIF at the crack tip decreases with $c$ thus the crack stops as soon
as the SIF is less than or equal to the toughness.  In cases of large
cracks and easy initiation, the scatter on toughness measurements is
thus considered as independent of the initiation step and directly
induced by the effect of the microstructural disorder on the crack
arrest. Note however that a delayed or late initiation step due for
instance to a very homogeneous material and defect free surface may be
responsible for an additional scatter (this is especially the case for
monocrystals where the crack propagation and arrest may
result from elasto-dynamic effects after an initiation step that
would allow for a large elastic energy being stored at the onset of
fracture). The present study only focuses on the effect of
microscopic disorder on the toughness measurements statistics in a
quasi-static regime.

Though not directly in the context of indentation, the recent
years have seen a growing interest for the study of fracture in
heterogeneous materials~\citep{Roux-Herrmann-book}. It appears
that many results obtained in the simple framework of the
propagation of a plane crack is applied in the more complex
case of the indentation geometry.  Historically, the apparent
universality of the scaling law characterizing the roughness of
crack surfaces~\citep{EBouchaud-JPC97} has motivated the use of
models initially developed in the Statistical Physics community.
The physics of a crack front arrested by an array of
obstacles~\citep{Gao-Rice-JAM89} is very similar to the one of a
triple contact line in wetting
experiments~\citep{Joanny-deGennes-JCP84} and is described as
a depinning transition~\citep{Schmittbuhl-PRL95}. In recent papers
such depinning models were used i) to estimate the dependence of
the macroscopic toughness on the details of the microscopic
disorder~\citep{RVH-EJMA03} and ii) to propose a material
independent description of the indentation crack arrest
statistics~\citep{CVHR-JMPS04}.

It appears in particular that a precise description of the
distribution of the toughness effectively ``seen'' by a front of
extension $L$ propagating through a heterogeneous material
characterized by a toughness disorder at the microscopic scale
$\xi$ is obtained. Independently of the details of the
microscopic disorder, this effective toughness distribution
$p(K_c,L/\xi)$ is characterized by a universal functional form
depending on two parameters only, namely, an asymptotic toughness
$K^*$ and the standard deviation  $\Sigma$ of an effective
toughness distribution
$$
p\left(K_c,\frac{L}{\xi}\right)=\psi\left(\frac{K^*-K_c}{\Sigma}\right)
$$
where $K^*$ is an intrinsic (constant) parameter of the material
while $\Sigma$ is size dependent,
$\Sigma=\Sigma_0(L/\xi)^{-1/\nu}$ depends only on the standard
deviation $\Sigma_0$ of the microscopic toughness distribution
and on the relative length of the front $L/\xi$, $\nu$ being here
a non trivial universal exponent~\citep{SVR-IJMPC02}. For crack
fronts of infinite extension, the distribution converges toward a
Dirac distribution at the macroscopic toughness value $K^*$.  For
all forms of microscopic toughness distribution, the statistical
distribution of the ``mesoscopic'' toughness will follow the
above universal form with a unique function $\psi$ as soon as
$L/\xi\gg 1$.  This result is adapted to the specific case of
indentation, and henceforth it will give rise to a universal form
of the crack arrest length distribution~\citep{CVHR-JMPS04}.

The aim of the present paper is to test the validity of the latter
predictions on a series of indentation experiments on various
brittle materials (glass, alumina, silicon carbide and silicon
nitride). The paper is organized as follows. The statistical
modeling of crack pinning is briefly introduced and adapted to
indentation. After presenting the experimental material and
methods, the experimental results concerning the statistics of
crack lengths at various load levels are analyzed within the
previous theoretical framework.

\section{Summary of the propagation model}

A semi-infinite mode-I crack in an infinite medium propagating
along a weak heterogeneous and virtual interface is considered.
The heterogeneity is modeled by a random field of local toughness,
with a small-scale correlation length $\xi$ above which
correlations are neglected.  As the system is loaded by
external forces up to the onset of crack propagation, the crack
front does not remain straight, but rather displays corrugations
so as to adjust the local SIF to the
facing toughness.  The difficulty of the problem essentially
arises from the coupling between the crack front morphology,
$h(x)$ and the local SIF, $K(x)$. The latter is taken into account
through the first order perturbation
computation~\citep{Gao-Rice-JAM89}
\be
K(x)=K_0\left(1+\frac{1}{\pi} \int_{-\infty}^{\infty}
\frac{h(x)-h(x')}{(x-x')^2} dx'\right)
\ee
where $K_0$ is the SIF that would be obtained in similar loading
conditions for a straight crack front.  Let us simply list here
the main results concerning this model as obtained in previous
studies~\citep{SVR-IJMPC02,RVH-EJMA03}.  For small disorder
amplitude, small system size, and/or long correlation length of
the toughness along the crack propagation direction, a regime of
``weak pinning'' is encountered where the crack front has a smooth
change, and the macroscopic toughness is simply equal to the
geometrical average of the local toughness.  More interestingly,
the most common situation, \ie the generic one for a large system
as compared to the heterogeneity scale, is a regime of
``{\it strong pinning}''.  The latter is a second order phase
transition where the control parameter is the macroscopic stress
intensity factor.  The critical point corresponds to the
macroscopic toughness below which crack propagation is arrested
and above which it is sustained forever. This criticality of the
crack propagation onset comes along with standard features of
second-order phase transitions such as the absence of
characteristic length scales, self-similarity, scaling with
universal critical exponents, a set of properties that are
exploited in the sequel.

One of the first characteristic features is the occurrence of a
correlated roughness of the crack front morphology with a long
range correlation. The absence of length scale imposes a
self-affine roughness where the front $h(x)$ remains statistically
invariant in the affine transformation $x\to \lambda x$; $h\to
\lambda^\zeta h$.  The roughness or Hurst exponent, $\zeta$, is
one critical exponent whose value --- determined from numerical
simulations --- amounts to $\zeta\approx
0.39$~\citep{Rosso-PRE02,VR-PRE04}. The global roughness
\be
  w(L)=(\langle h(x)^2\rangle-\langle h(x)\rangle_L^2)^{1/2}
\ee
therefore scales with the system size as $w(L)\propto L^\zeta$.  Note
that experimental roughness measurements on interfacial propagating
fronts were shown to exhibit a self-affine character with a roughness
exponent $\zeta\approx0.5-0.6$ \cite{Schmittbuhl-PRL97} larger than
the above numerical value $\zeta\approx 0.39$. In the following we use
the latter  value in order to test te validity of our model
on experimental data.

The statistical distribution of {\it instantaneous} depinning
force is non-universal.  It reflects both the crack front morphology and
the  statistical distribution of local toughness. However, the
depinning is organized in space and time in such a way that the
front advance proceeds by a series of unstable jumps from one
stable configuration to the next with a large (scale free, and
hence power law) statistical distribution of such avalanches. The
latter displays a hierarchical structure consisting of embedded
sub-avalanches with a self-similar statistical distribution.  The
interesting feature is that the statistical distribution of depinning
forces or toughness $K_c$, $p(K_c,\ell)$, conditioned by the
avalanche size (or more precisely by a characteristic distance
$\ell$ along the crack front) obeys some important scaling
properties:
  \begin{itemize}
  \item
The most important, is that as $\ell$ tends to infinity, the
distribution tends to a Dirac distribution
  \be
  p(K_c,\ell)\longrightarrow \delta(K_c-K^*)
  \ee
where $K^*$ is the thermodynamic limit of the toughness.  It is
thus retrieved that for an infinite system size (relative to the
size of the heterogeneities along the interface) the system
converges toward a homogeneous one characterized by a
deterministic toughness $K^*$.

\item
The way the convergence toward this deterministic limit occurs
is further characterized.  For a finite characteristic length
scale $\ell$, the distribution $p(K_c,\ell)$ has a standard
deviation $\Sigma(\ell)=(\langle K_c^2\rangle-\langle K_c\rangle^2
)^{1/2}$ that vanishes for diverging $\ell$ as
  \be
  \Sigma(\ell)\propto \Sigma_0 \left(\frac{\ell}{\xi}\right)^{-1/\nu}
  \ee
where $\nu=1/(1-\zeta)\approx 1.64$ is a universal exponent.
Moreover, these two moments are sufficient to characterize the
entire force distribution.

\item{}
The reduced toughness defined as $u=(K^*-K_c)(\ell/\xi)^{1/\nu}$
is observed (for large $\ell$) to follow a universal distribution
  \be
  p(K_c,\ell)=\left( \frac{\ell}{\xi}\right)^{1/\nu}\psi(u)
  \ee
where $\psi$ is universal, that is $\ell$-independent, but more
importantly also independent of the details of the local toughness
distribution. In the above definition, $\psi$ is defined at the
scale $\xi$ of the microscopic disorder, its shape is universal
and its variance is equal to $\Sigma_0^2$, which is the only
relevant microscopic parameter. For small arguments $u\ll 1$,
$\psi(u)$ behaves as a power law
 \be
\psi(u) \approx A_0\frac{u^\beta}{\Sigma_0^{\beta+1}}
\ee
where $A_0$ is a constant of the order of unity and
$\beta=\zeta/(1-\zeta)\approx 0.64$ is again a critical exponent.
For large arguments $\psi$ decays sharply to 0 (\ie faster than
any power law).
\item{}
However, the record in time of the macroscopic toughness displays
long range time correlations, which forbid all practical use of
the above result without resorting to a more detailed study. These
correlations are however exhausted past a characteristic
propagation distance that exactly matches the front roughness. The
condition for negligible correlations in the crack propagation is
that the crack front has been renewed over its entire extension.
Thus for a system size $L$, the distribution $p(K_c,\ell=L)$ both
displays the universal shape given by $\psi$, and the absence of
correlations along the crack propagation axis which makes the
information exploitable and useful.
\end{itemize}

It is highly non trivial that only two parameters $K^*$ and
$\Sigma$ (for a reference scale $L$) are sufficient to account
for the entire distribution for any statistical distribution of
local toughness. Using cautiously a loose analogy, this behavior
is reminiscent of the central limit theorem, concerning the
distribution of the average $S_N$ of $N$ random numbers $S_N=1/N
\sum x_i$. The macroscopic toughness limit, $K^*$, would play the
role as $\langle x\rangle$, the variance of $S_N$ decreases as a
power-law of $N$, as the variance of the toughness does with the
crack length, and the rescaled distribution of $(S_N-\langle
S_N\rangle)/\sqrt{\langle S_N^2\rangle-\langle S_N\rangle^2}$
follows a universal (here Gaussian) distribution comparable to
$\psi$ in the sense of its independence with respect to the
distribution of $x_i$ or local toughness. In the following, this
scaling property is exploited in the indentation geometry to
develop a unified analysis of crack arrest statistics.

\section{Application to indentation crack arrest}

A very simple approach to crack arrest statistics is developed
within a one-dimensional picture. Assuming a crack propagating
across a layered material, the crack length statistics  is
written as~\citep{1577,1272,1252,1322}
\be
Q(c)=\prod_{i=1}^{c/\xi} F\left[K(i\xi)\right] \approx \exp
\left(\frac{1}{\xi}\int_0^c \log F[K(x)] dx \right)
\ee
where $Q(c)$ is the probability of having a crack length greater
than $c$, $F[K_c]$ the probability that the toughness be less than
$K_c$, $K(x)$ the value of the SIF at location $x$ and $\xi$ is
the width of one layer or the correlation length in case of a
random continuous toughness field. This result simply relies on
the statistical independence of the toughness value between
successive layers. It also allows one to account for the
dependence of the SIF on the crack length.  However, as such, it
is dependent on the material properties through the entire
function $F$, and the interpretation of the length scale $\xi$
although clear for a layered system remains to be clarified for
more heterogeneous systems.

Let us now address the question of crack arrest in a Vickers
indentation experiment, considering a radial crack system.  From
the above section, some fundamental results are at hand to reduce
significantly the specificity of the problem applied to a given
material. One particularity of the problem is that the (radial)
crack is semi-circular of radius $c$ and hence the crack length
$L$ is proportional to its propagation distance $c$. In the
modeling of the problem as one-dimensional (\ie simply
parameterized by $c$), to avoid correlations in the global
critical SIF one has to resort to a coarse-grained discrete
description. The crack front is correlated over its entire length
$L=\pi c$ and over a width $w\propto \sigma_0
\xi^{1-\zeta}L^\zeta$ where $\sigma_0=\Sigma_0/K^*$ corresponds
to the standard deviation of the relative toughness fluctuations
at the microscopic scale $\xi$. In the indentation geometry, the
crack is thus regarded as propagating through a series of
discrete shells whose width depends on the radius as $w\propto
c^\zeta$.  Moreover, each of these shells, will have a different
statistical distribution of toughness (same $K^*$ but different
width $c^{-1/\nu}$). Yet, these unexpected features are taken
into account rigorously, and one arrives (see~\citep{CVHR-JMPS04}
for details) at the full expression of the crack length
probability distribution.

Let us first introduce $c^*$, the crack radius which would occur
if the toughness were homogeneous at its thermodynamic asymptotic
value $K^*$, \ie $K(c^*)=K^*$. As mentioned earlier, in the
present case of indentation, the toughness disorder induces a
systematic $R$-curve effect. The larger the applied mass, the
larger the apparent toughness (although the upper bound does not
change with the crack size, whereas the mean value does). For a
given applied mass, the value $c^*$ is a lower bound on the crack
size at arrest.  For a large crack size as compared to the scale
of heterogeneities, the propagation probability is written as
\begin{eqnarray}
 \label{eq:2Db}
 Q(c) & \approx &
 \exp\left\{-A_0~\sigma_0^{\frac{\zeta-2}{1-\zeta}}
 \left(\frac{c^*}{\xi}\right)^{2-\zeta}\right. \nonumber \\
   & \times & \left. \int_{c^*}^c
   \left(\frac{r}{c^*}\right)^{1-\zeta}
   \left(\frac{K^*-K(r)}{K^*}\right)^{\frac{1}{1-\zeta}}
 \frac{\mathrm{d}r}{c^*} \right\}
\end{eqnarray}
where again $A_0$ is a constant of the order of unity.

Let us further assume  that the SIF follows a power law decrease
with the crack radius $K(c)\propto F/c^{m}$, where asymptotically
(for large loads) $m=3/2$.  In fact, one may argue that this
essentially elastic-brittle description ($m=3/2$) has to be
corrected by a dimensionless function of the ratio $c/a$ to
account for plasticity effects for $c/a<2$.  Lacking a safe
experimental background for justifying such a systematic effect,
an effective general power law dependence is introduced, and
reverts to the value $m=3/2$ for large loads or as a first
estimate for $m$
\begin{eqnarray}
   \label{eq:2Dc}
   Q(c) & \approx &
   \exp\left\{ -A_0~ \sigma_0^{\frac{\zeta-2}{1-\zeta}}
   \left(\frac{c^*}{\xi}\right)^{2-\zeta}\right. \nonumber \\
   & \times & \left. \int_1^{x}
   u^{1-\zeta} \left(1-u^{-m}\right)^{1/(1-\zeta)}
   \mathrm{d}u \right\}
\end{eqnarray}
with $x=c/c^*$. The integral in Eq.~(\ref{eq:2Dc}) is
recast in a more compact form
%terms of an incomplete beta function
%
\be
 \label{eq:2Dcn}
 Q(c)=\exp\left[-A_0\left(\frac{c^*/\xi}{\sigma_0^{1/1-\zeta}}\right)^{2-\zeta}
 \mathcal{B}\left(\frac{\zeta-2}{m},\frac{2-\zeta}{1-\zeta},
 \left(\frac{c}{c^*}\right)^{-m}\right)
 \right]
\ee
where $\mathcal{B}$ is based on an incomplete beta function
\be
\label{beta}
    \mathcal{B}(\mu,\eta,x) = \frac{\mu(1-\eta)}{\eta}
    \int_{x}^1 \tau^{\mu-1}(1-\tau)^{\eta-1}
    \mathrm{d}\tau
\ee
The width of the distribution thus only depends i) on the
system size $c^*/\xi$ and ii) on the strength $\sigma_0$ of the
relative toughness disorder at the microscopic scale.

It may be noted that the same type of distribution was chosen {\em
a priori} by~\citet{1252}. Equation~(\ref{eq:2Dc}) therefore
constitutes an {\em a posteriori} validation, even though the
``grain size'' is no longer constant in the present analysis. Let
us also recall that $2-\zeta \approx
1.61$~\citep{Rosso-PRE02,VR-PRE04}.

\section{Experimental material and methods}

The previous analytical results are now tested against
experimental data. Several indentation tests, with different
applied masses $M$ (\ie $F \propto M$), have been performed ($M=$
0.2~kg, 0.3~kg, 0.5~kg and 1~kg) on four different brittle
materials, namely alumina (Al$_2$O$_3$), silicon nitride
(Si$_3$N$_4$), silicon carbide (SiC) and sodalime silicate glass.
The microstructure of alumina is made of fine-grain (\ie 10$\mu$m)
alumina polycrystals with an inter-granular glassy phase.  A
microanalysis shows that the latter contains SiO$_2$, CaO and
Al$_2$O$_3$ components. Silicon nitride is isostatically pressed
with an average grain size close to 3~$\mu$m. The silicon carbide
material is sintered. The powder was pressed and heated to 2000
degrees Celsius. During this step, small quantities of boron
carbide have been added to improve the sintering process. A small
porosity is induced by this manufacturing process. The SiC
microstructure consists of grains whose characteristic length is
estimated to 4~$\mu$m. Last, the studied glass is a standard float
grade, of typical composition 72~wt$\%$ SiO$_2$, 14~wt$\%$
Na$_2$O, 9.5~wt$\%$ CaO, 4.5~wt$\%$ MgO, with traces of K$_2$O,
Fe$_2$0$_3$, and Al$_2$O$_3$. Beyond the nanometer scale glass is
regarded as homogeneous.

After each indentation test, the length $c$ of the cracks and the
diagonal $2a$ of the plastic print are measured. For the chosen
loading range, it was checked that the radial-median crack system
is predominant, following a $c/a$ criterion~\citep{1241,1242}. In
the case of alumina, additional observations have been performed
to check that indentation-generated cracks (for an applied mass of
1 kg) remained connected to the plastic mark after polishing,
thus discriminating the radial-median crack system from a
Palmqvist one. Consequently only indentation results related to a
ratio $c/2a$ greater than 1 are used with the only exception of
results obtained on Si$_3$N$_4$ for which all crack lengths are
such that $c/2a<1$. In the latter case, the Palmqvist crack system
is likely to be predominant. The questions of the identification
and the effect of the crack system on the statistical analysis
presented in this paper are discussed in Section~\ref{palmqvist}
in more details. For each series of tests performed with the same
load, the measured crack lengths $c_i$ are associated to an
experimental propagation probability $Q(c_i)$. The crack lengths
are ranked in ascending order (\ie $c_1 < c_2 < \ldots < c_N$) and
the corresponding experimental probability is evaluated as $Q(c_i)
= 1 - i / (N+1)$, where $N$ is the number of measured crack radii
for the considered applied mass.

Before any indentation test, the given material sample has been
polished. This allows one to present an almost perfectly planar
surface below the indentor and to better control the
perpendicularity of the surface with respect to the indentation
axis. Furthermore, this additional polishing process substantially
decreases possible surface residual stresses which would affect
the crack propagation conditions, and introduce some bias in the
crack length measurements.

Beyond uncertainties due to the sample preparation, the measured
indentation-generated crack length is strongly dependent on both
the observation equipment and the experimentalist. To account for
these parameters, one may compare data obtained with two different
indentors on the same material and under the same conditions. It
appears that the measured crack length strongly depends on the
optical observation, but its influence on the measured crack
length {\em scatter} is rather low. Last, one compares results
obtained by two different operators measuring the same crack
length, in order to evaluate its influence in the definition of
the crack tip location. Again, although the measured absolute
crack length may be operator-dependent, the scatter remains
comparable.

\section{Identification results}

In the following analysis, data obtained on a sodalime silicate
glass are analyzed to discuss different identification strategies.
Equation~(\ref{eq:2Dc}) is simply rewritten as
\be
    \label{eq:2Dcnb}
    Q(c)=\exp\left[-A~(c^*)^{2-\zeta}
    \mathcal{B}\left(\frac{\zeta-2}{m},\frac{2-\zeta}{1-\zeta},
    \left(\frac{c}{c^*}\right)^{-m}\right)
    \right]
\ee
where
\be
    A=A_0\left(\xi \sigma_0^{1/1-\zeta} \right)^{\zeta-2}
\ee
Only two parameters are to be identified, namely the scale
parameter $A$ and the characteristic radius $c^*$ provided the
value for $m$ is known. A least squares technique is used to
determine the unknown parameters by minimizing the difference
between the measured and modeled propagation probabilities.
Furthermore, a rescaling procedure is followed to collapse all the
experimental data onto a {\em single} master curve. The rescaled
propagation probability $\tilde{Q}$ is defined as
\be
    \tilde{Q} = Q^q
\ee
and the dimensionless crack radius $\tilde{c}$
\be
    \tilde{c} = \frac{c}{c^*}
\ee
where $q$ will depend on the type of identification procedure.
Different strategies are followed to analyze the experimental
data.

\subsection{Step 1: Test of the distribution shape}

\begin{figure}[ht!]
%\centerline{\epsfxsize=.9\hsize \epsffile{verre/ident_new1_glass.eps}}
%\centerline{\scalebox{0.7}{\includegraphics{verre/ident_new1_glass.eps}}}
\centerline{\epsfig{file=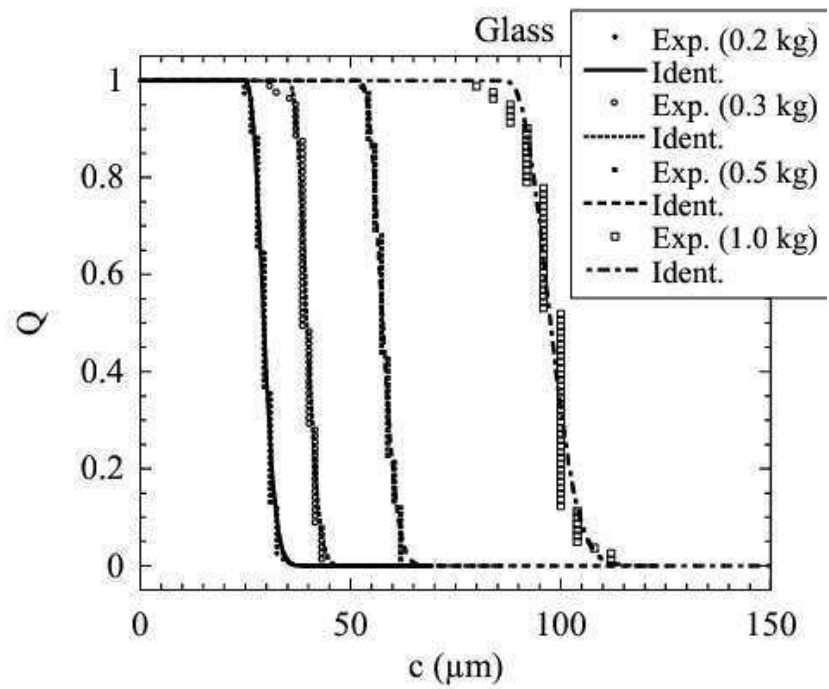,width=0.95\hsize}}
    \vfill
\caption{Propagation probability $Q$ versus crack radius $c$ for
four different applied masses on sodalime silicate glass. The
symbols are experimental data and the lines are identifications
when each load level is analyzed independently.}
\label{fi:ident_new1_glass}
\end{figure}

\begin{figure}[ht!]
%\centerline{\epsfxsize=.9\hsize
%\epsffile{verre/parameters_glass.eps}}
%\centerline{\scalebox{0.7}{\includegraphics{verre/parameters_glass.eps}}}
\centerline{\epsfig{file=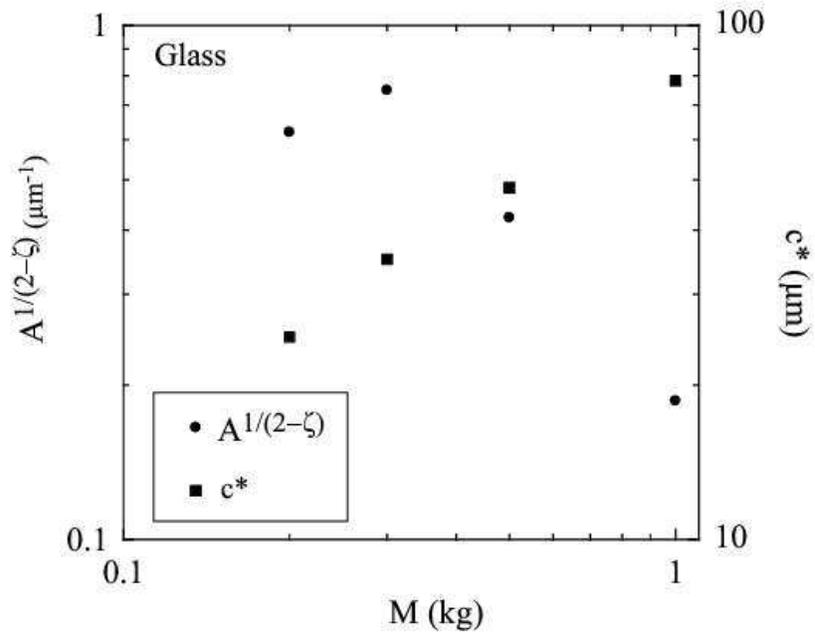,width=0.95\hsize}}
    \vfill
\caption{Parameters $A$ and $c^*$ versus applied mass $M$ when
each load level is analyzed independently for sodalime silicate
glass.} \label{fi:parameters_glass}
\end{figure}

\begin{figure}[ht!]
%\centerline{\epsfxsize=.9\hsize
%\epsffile{verre/ident_new1_col_glass.eps}}
%\centerline{\scalebox{0.7}{\includegraphics{verre/ident_new1_col_glass.eps}}}
\centerline{\epsfig{file=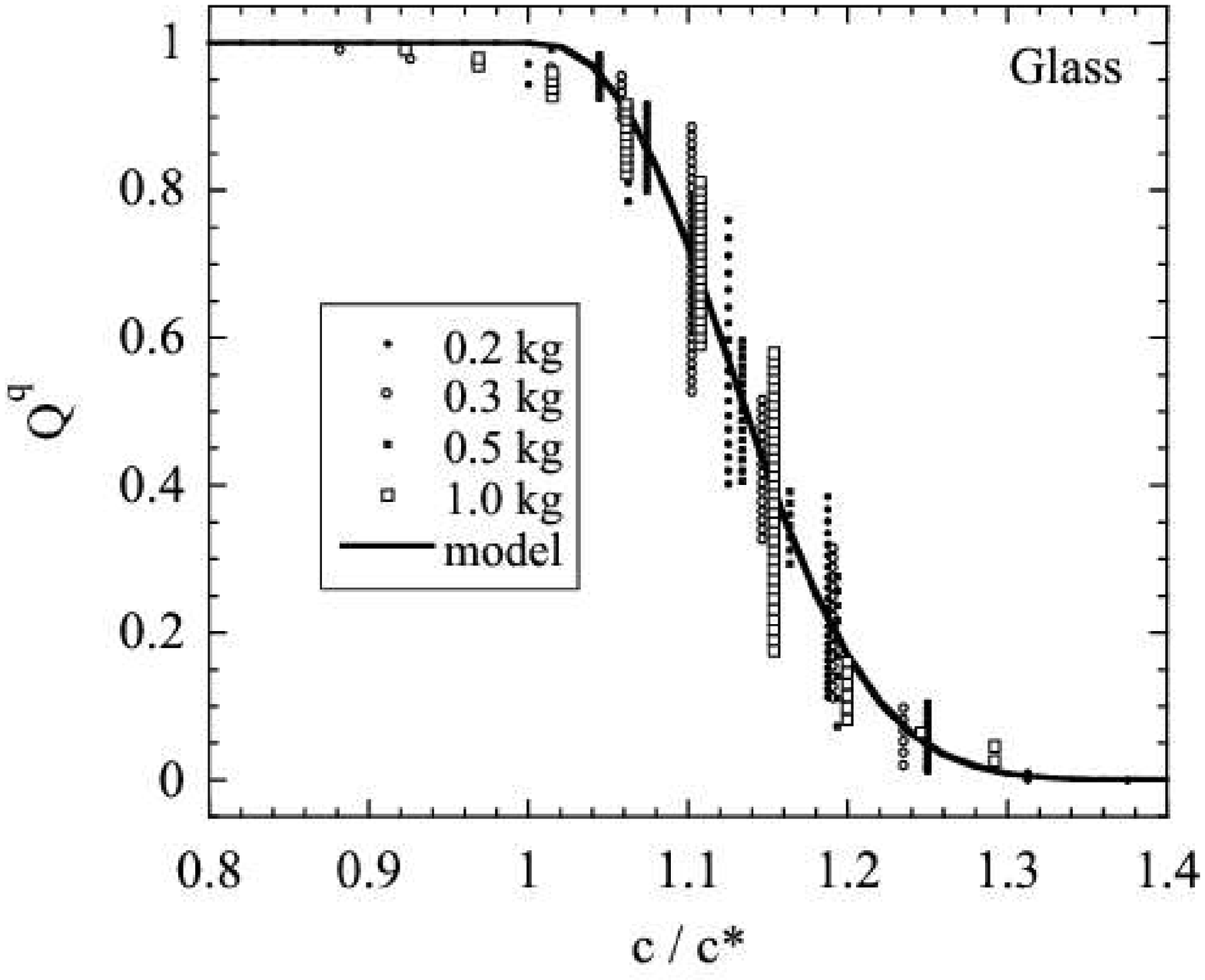,width=0.95\hsize}}

\caption{Rescaled propagation probability $Q^q$ versus
dimensionless crack radius $c/c^*$ for four different applied
masses. The symbols are experimental data of sodalime silicate
glass and the line is the result of the identification. From the
present analysis, it is expected that all experimental points
should fall on the same curve.} \label{fi:ident_new1_col_glass}
\end{figure}

The first test concerns the ability of the proposed scaling form
to account for the shape of the arrest length distribution, for
each individual material and load level {\em treated
independently}. For this purpose, the $m$ exponent is set to its
asymptotic value $m=3/2$. Figure~\ref{fi:ident_new1_glass} shows
experimental data obtained for sodalime silicate glass with the
four different masses and the corresponding identification for
each load. A good agreement is obtained in terms of overall
distribution, thereby validating the general form of the
distribution given in Eq.~(\ref{eq:2Dc}). Both parameters depend
upon the applied mass as shown in Fig.~\ref{fi:parameters_glass}.
Figure~\ref{fi:ident_new1_col_glass} shows the result of the
rescaling procedure. In this approach, the power $q$ is defined as
\be
    q = \frac{B_0}{A(c^*)^{2-\zeta}}
\ee
in which $B_0$ is chosen as the geometric average of the products
$A(c^*)^{2-\zeta}$ for all the applied masses. It is noted that
out of 458 measurement points, only 18 are such that $c/c^* < 1$ (none
is expected theoretically).

\subsection{Step 2: Test of the objectivity of material parameters}

In the next identification stage, it is assumed that the parameter
$A$ is load-independent, as expected from Eq.~(\ref{eq:2Dc}), and
characteristic of the material.  Thus a further check of the
prediction is that a single value for all load levels should
account for all distributions. The constancy of $A$ is thus
prescribed during the identification stage. In contrast, the
radius $c^*$ is still assumed to be load-dependent.  It is to be
noted that $c^*$ is the result of an intrinsic material parameter
$K^*$, together with a relationship between $K$, $F$ and the crack
radius modeled by a power law of exponent $m$.  As one dealt
experimentally with rather short cracks, the latter relation may
deviate from the simple power law dependence, $m=3/2$.  Thus at
this stage, $m$ is still set to its asymptotic value, but $c^*$ is
considered as a free parameter determined for each load.

\begin{figure}[ht!]
%\centerline{\epsfxsize=.9\hsize
%\epsffile{verre/ident_bis_new1_glass.eps}}
%\centerline{\scalebox{0.7}{\includegraphics{verre/ident_bis_new1_glass.eps}}}

\centerline{\epsfig{file=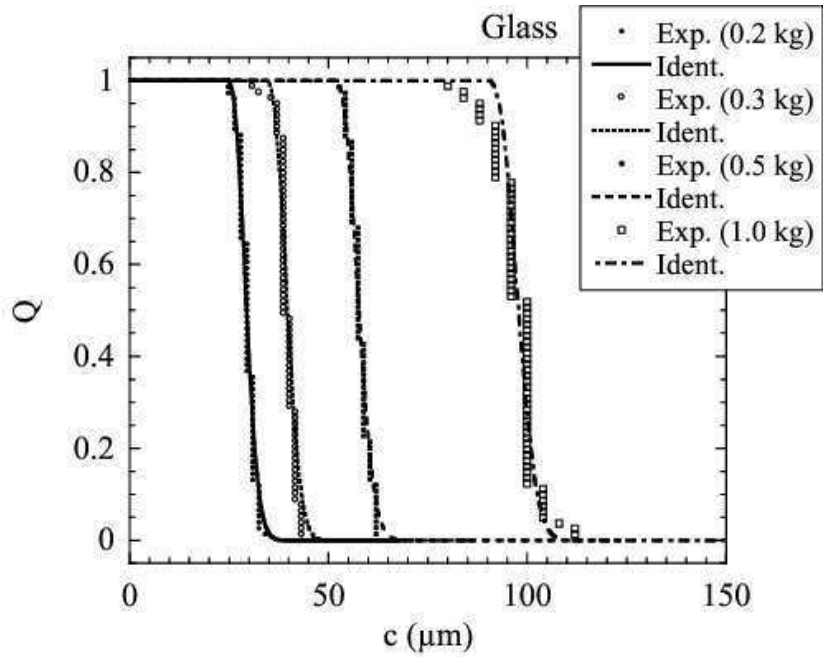,width=0.95\hsize}}

\caption{Propagation probability $Q$ versus crack radius $c$ for
four different applied masses on sodalime silicate glass. The
symbols are experimental data and the lines are identifications
when the parameter $A$ is assumed to be load-independent.}
\label{fi:ident_bis_new1_glass}
\end{figure}

\begin{figure}[ht!]
%\centerline{\epsfxsize=.9\hsize
%\epsffile{verre/ident_bis_new1_col_glass.eps}}
%\centerline{\scalebox{0.7}{\includegraphics{verre/ident_bis_new1_col_glass.eps}}}
\centerline{\epsfig{file=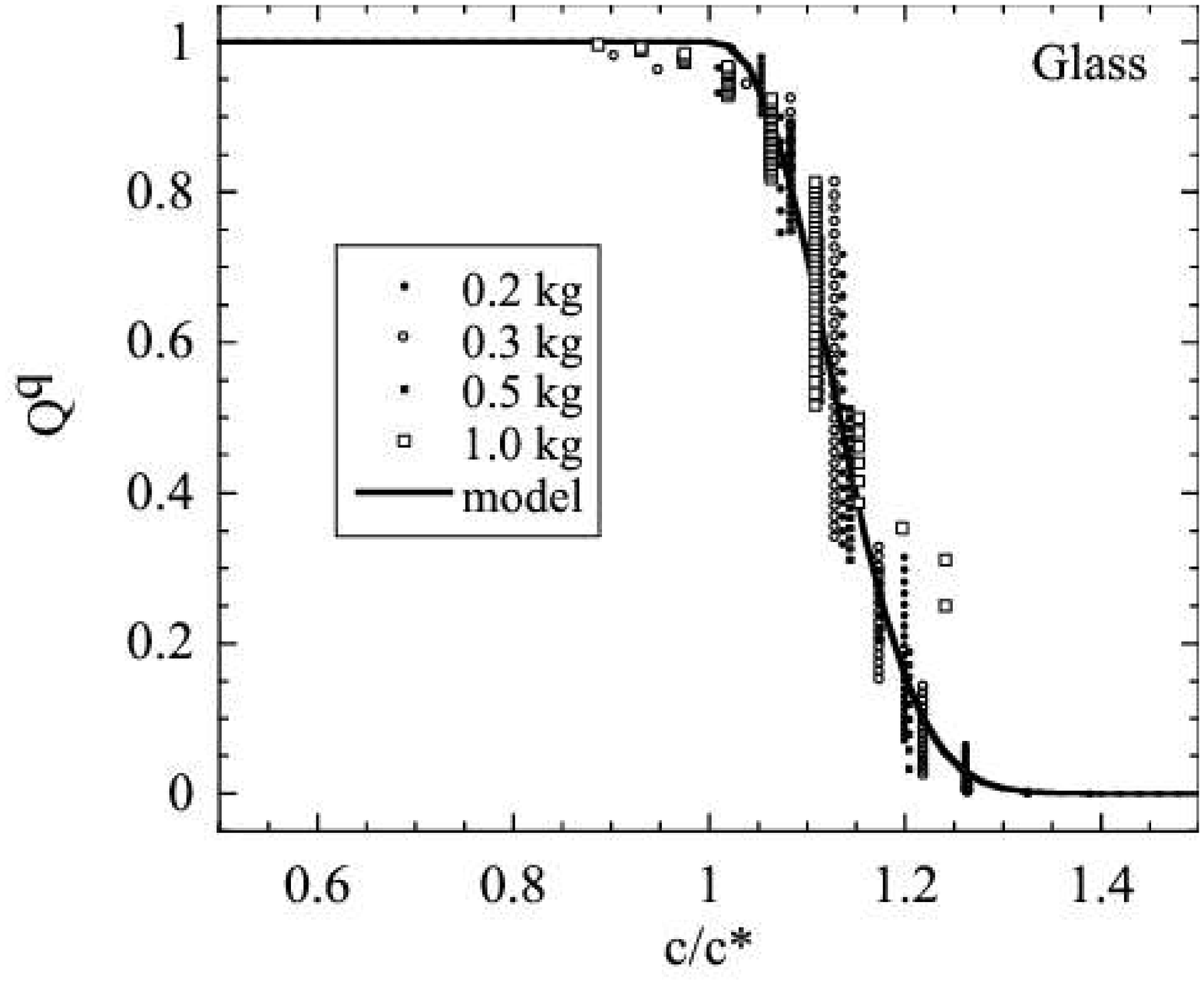,width=0.95\hsize}}
\caption{Rescaled propagation probability $Q^q$ versus
dimensionless crack radius $c/c^*$ for four different applied
masses. The symbols are experimental data on sodalime silicate
glass and the line is the result of the identification with a
constant parameter $A$. From the present analysis, it is expected
that all experimental points should fall on the same curve.}
\label{fi:ident_bis_new1_col_glass}
\end{figure}

Figure~\ref{fi:ident_bis_new1_glass} shows the result of the
identification. A good agreement is obtained, even though not as
good as in the previous case (as naturally expected because the
functional form is more constrained). The value of
$A^{1/(2-\zeta)}$ is found to be equal to 0.57~$\mu$m$^{-1}$. Let
us recall that this value is not directly read as an (inverse)
characteristic correlation length that would signal the physical
size of heterogeneities. The amplitude of the microscopic
toughness distribution, $\sigma_0$ does contribute to $A$.
Figure~\ref{fi:cstar-m_glass} shows the dependence of the
characteristic radius $c^*$ with the applied mass $M$. The
observed dependence is close to what is expected from the model
[\ie $c^* \propto M^{2/3}$ through the definition of a stress
intensity factor $K^* \propto M / (c^*)^{3/2}$]. Since the
parameter $A$ is constant, the power $q$ of the rescaled
propagation probability is now given by
\be
    q = \left(\frac{c_0}{c^*}\right)^{2-\zeta}
\ee
in which $c_0$ is chosen as the geometric average of the
characteristic radii $c^*$ for all the applied masses.
Figure~\ref{fi:ident_bis_new1_col_glass} shows the prediction for
the indentation experiments on glass. Only nine points are such
that $c/c^* < 1$.

\subsection{Step 3: Complete test of the prediction}

\begin{figure}[ht!]
%\centerline{\epsfxsize=.9\hsize
%\epsffile{verre/cstar-m_glass.eps}}
%\centerline{\scalebox{0.7}{\includegraphics{verre/cstar-m_glass.eps}}}
\centerline{\epsfig{file=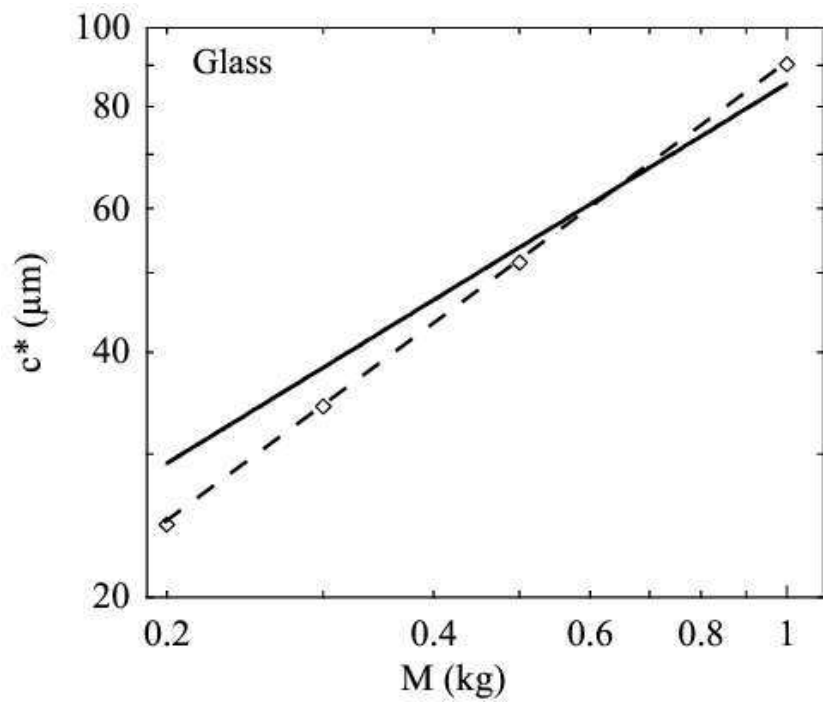,width=0.95\hsize}}
\caption{Parameter $c^*$ versus applied mass $M$. A power law with
an exponent of 2/3 fits reasonably the experiments on sodalime
silicate glass (solid line). The dashed line corresponds to the
best power law fit for an exponent equal to 0.81.}
\label{fi:cstar-m_glass}
\end{figure}

The final test would consist in enforcing the $K(c)$ dependence
with $m=3/2$ in the identification procedure. It leads to a
significant degradation of the quality of the results. As
mentioned earlier, in most of the test cases, the ratio of the
crack length to the half diagonal of the plastic imprint is too
small to trust the asymptotic $K(c)$ law.  As observed in the
previous step, Fig.~\ref{fi:cstar-m_glass}, $c^*$ deviates from
the expected power law of exponent $1/m=2/3$. Such deviations are
not surprising when dealing with small values of the ratio $c/a$,
plasticity being thus the main dissipation mechanism.

However, one also notes that an {\em effective} power law fits
the data quite nicely.  Let us insist on the fact that this is a
purely empirical observation, which is not supported by any
theoretical argument, nor by the literature that proposes numerous
conflicting laws~\citep{1241}. If one introduces $m'$ the exponent
such that
\be
 c^*\propto M^{1/m'}
\ee
one measures $1/m'\approx 0.8$ for glass
(Fig.~\ref{fi:cstar-m_glass}). Note that the scaling $K\propto
M/c^{3/2}$ is only expected to hold for $c/a>2$; $c/a$ dependent
corrections to scaling being necessary when $c/2a$ is closer to
unity. The value of the effective exponent $1/m$ is thus expected
to approach  $2/3$ as the ratio $c/a$ increases. See also
Subsection~\ref{palmqvist} for a detailed discussion about the
dependence of $m$ on the nature of the crack system.

Such an observation is however unsatisfactory in the sense that a
power law dependence involving $m$ has already been used in the
derivation of the functional forms which allowed one to estimate $m'$.
Thus, as a final self-consistency requirement, it is proposed to
determine the best parameter $m$ such that the observed $m'$
matches its starting value $m$. This self-consistent $m$ value is
determined numerically using a fixed point algorithm, \ie setting
$m$ to a previously determined $m'$ until convergence. In one
iteration the value of the power is unaltered for three
representative digits. Furthermore, the predictions are very close
to those obtained in Figs.~\ref{fi:ident_bis_new1_glass} and
\ref{fi:ident_bis_new1_col_glass} since the
initial value $m'$ is already very close (\ie 99~\%) to its
converged estimate.

\subsection{Application to four different brittle materials}

The results obtained for the four different brittle materials
(Al$_2$O$_3$, SiC, Si$_3$N$_4$, glass) are presented when using
the above described statistical treatments.

\subsubsection{Step 1}

\begin{figure}[ht!]
%\centerline{\epsfxsize=.6\hsize
%\epsffile{Al2O3/ident_new1_col_al2o3.eps}}
%\centerline{\scalebox{0.45}{\includegraphics{Al2O3/ident_new1_col_al2o3.eps}}}
\centerline{\epsfig{file=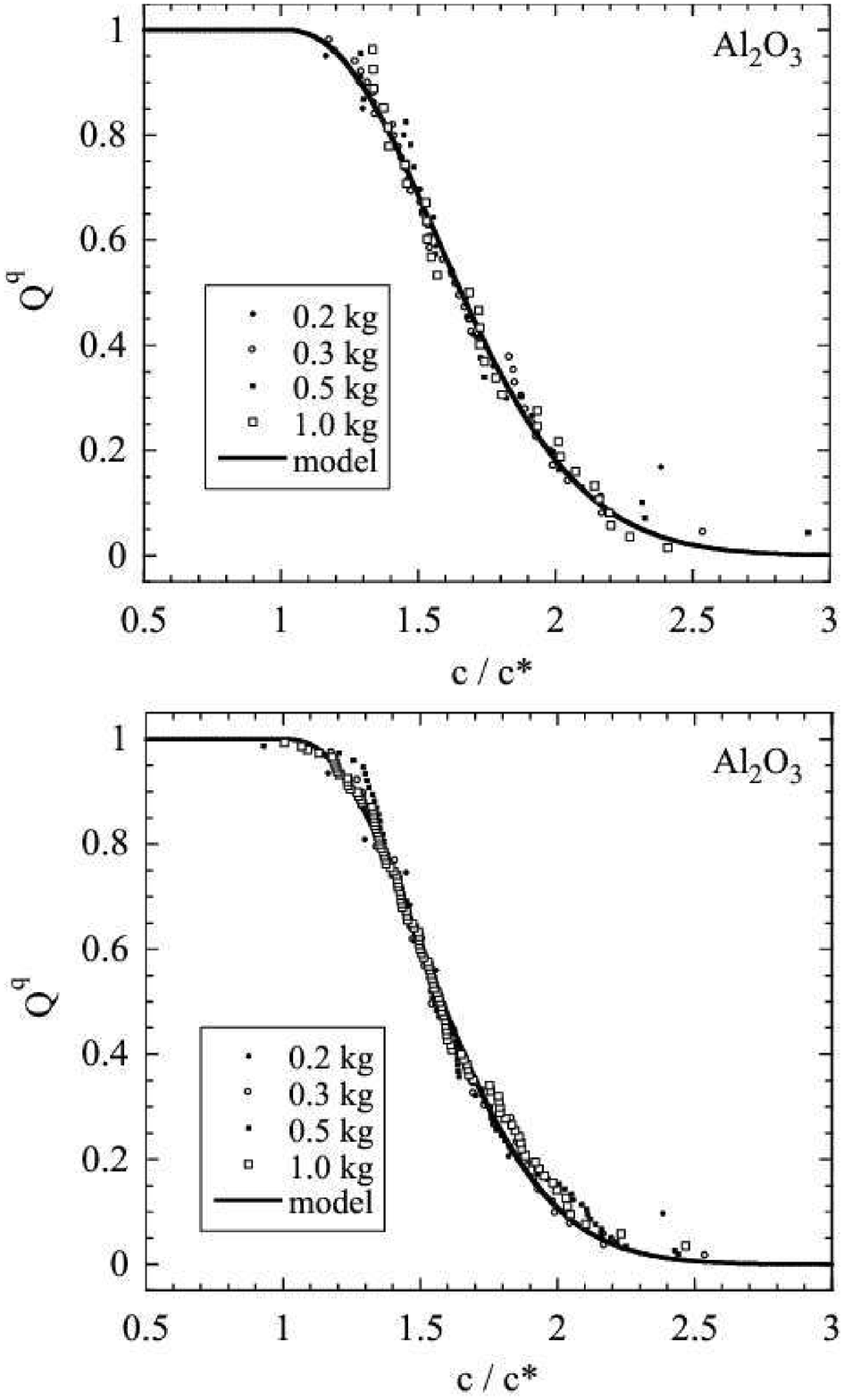,width=0.95\hsize}}
\caption{Rescaled propagation probability $Q^q$ versus
dimensionless crack radius $c/c^*$ for four different applied
masses on alumina. The symbols are experimental data and the line
is the result of the identification. From the present analysis, it
is expected that all experimental points should fall on the same
curve. Top: 1 machine and 1 operator. Bottom: 2 indentors and 2
operators.} \label{fi:ident_new1_col_al2o3}
\end{figure}

\begin{figure}[ht!]
%\centerline{\epsfxsize=.9\hsize
%\epsffile{Si3N4/ident_new1_col_si3n4.eps}}
%\centerline{\scalebox{0.7}{\includegraphics{Si3N4/ident_new1_col_si3n4.eps}}}
\centerline{\epsfig{file=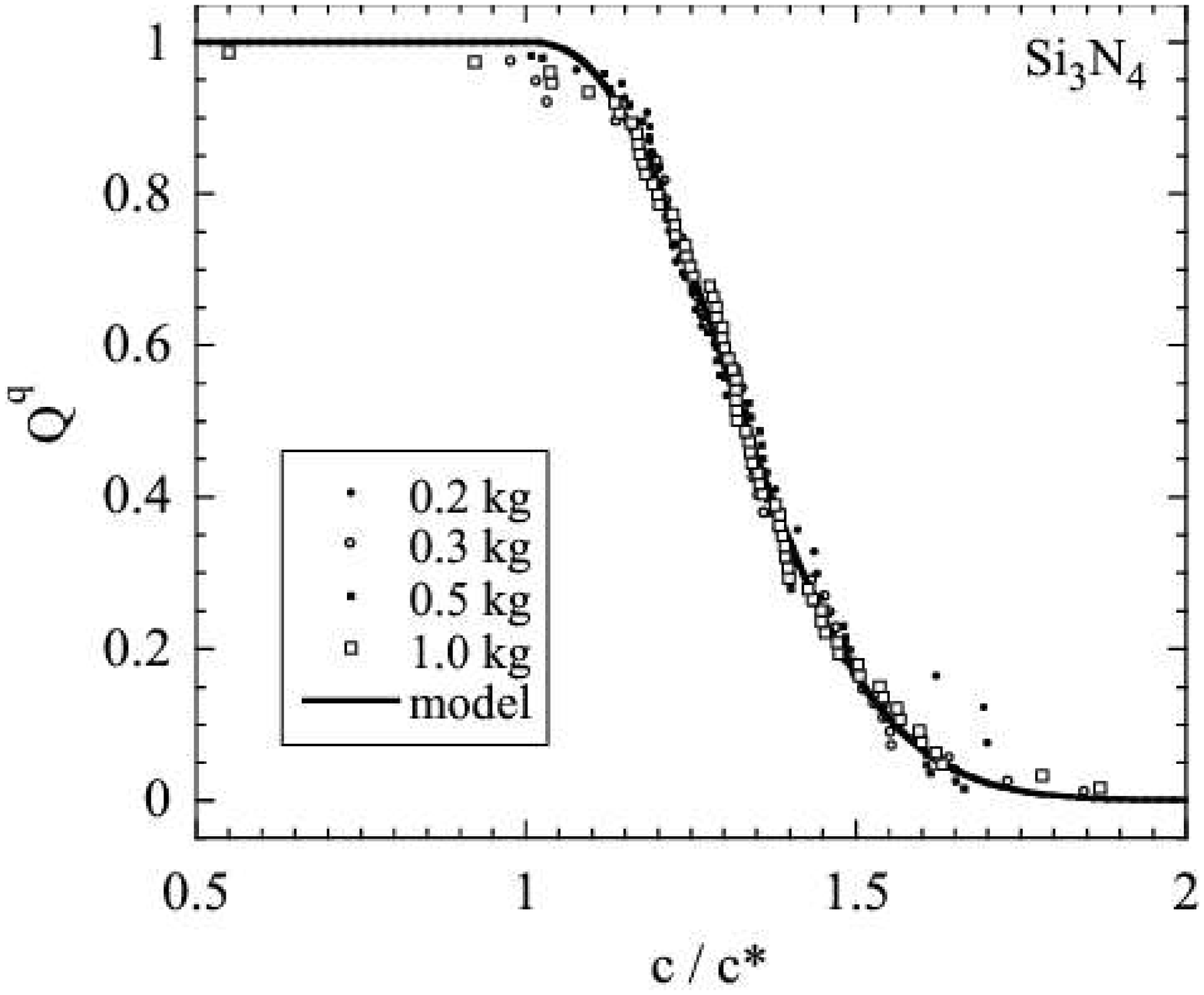,width=0.95\hsize}}
\caption{Rescaled propagation probability $Q^q$ versus
dimensionless crack radius $c/c^*$ for four different applied
masses on silicon nitride. The symbols are experimental data and
the line is the result of the identification. From the present
analysis, it is expected that all experimental points should fall
onto the same curve.} \label{fi:ident_new1_col_si3n4}
\end{figure}

\begin{figure}[ht!]
%\centerline{\epsfxsize=.6\hsize
%\epsffile{SiC/ident_new1_col_sic.eps}}
%\centerline{\scalebox{0.45}{\includegraphics{SiC/ident_new1_col_sic.eps}}}
\centerline{\epsfig{file=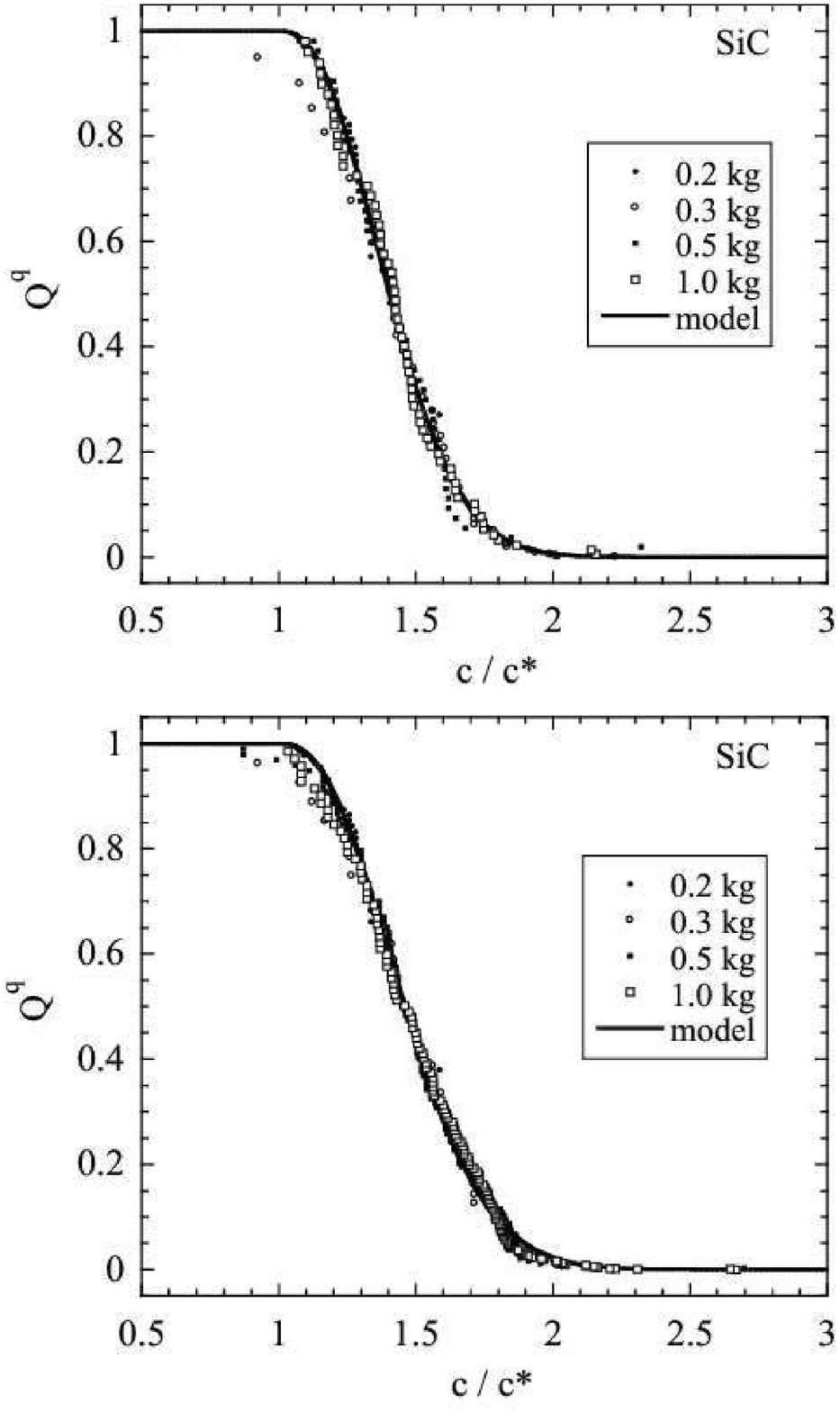,width=0.95\hsize}}

\caption{Rescaled propagation probability $Q^q$ versus
dimensionless crack radius $c/c^*$ for four different applied
masses on silicon carbide. The symbols are experimental data and
the line is the result of the identification. From the present
analysis, it is expected that all experimental points should fall
on the same curve. Top: 1 machine and 1 operator. Bottom: 2
indentors and 2 operators.} \label{fi:ident_new1_col_sic}
\end{figure}

\begin{figure}[ht!]
%\label{fi:ident_new1_col_glass_all}
%\centerline{\epsfxsize=.6\hsize
%\epsffile{verre/ident_new1_col_glass_all.eps}}
%\centerline{\scalebox{0.45}{\includegraphics{verre/ident_new1_col_glass_all.eps}}}
 \centerline{\epsfig{file=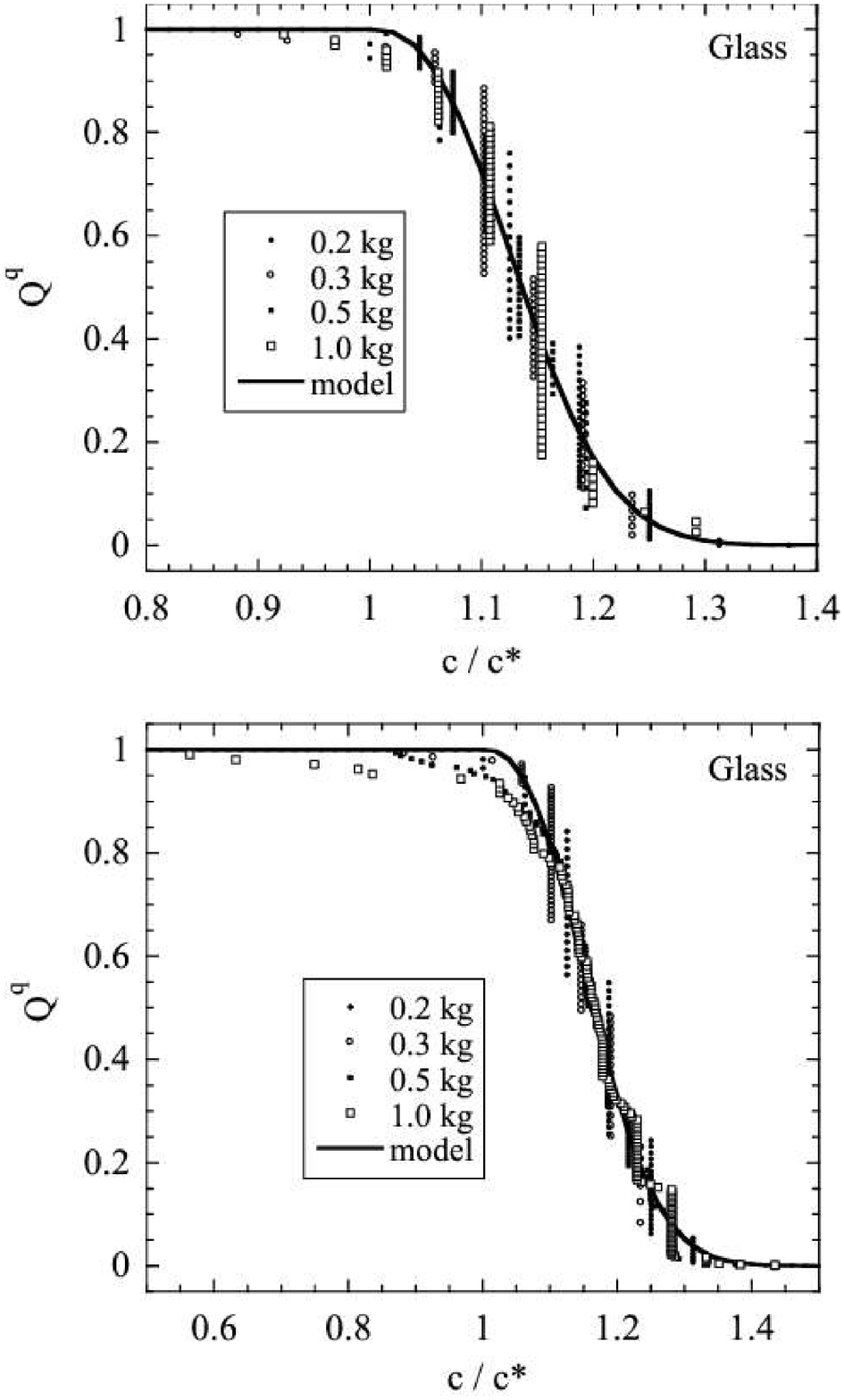,width=0.95\hsize}}

\caption{Rescaled propagation probability $Q^q$ versus
dimensionless crack radius $c/c^*$ for four different applied
masses on sodalime silicate glass. The symbols are experimental data and the line is
the result of the identification. From the present analysis, it is
expected that all experimental points should fall on the same
curve. Top: 1 machine and 1 operator. Bottom: 2 indentors and 2
operators.} \label{fi:ident_new1_col_glass_all}
\end{figure}

By following the first identification stage, one obtains the
following results:

\begin{itemize}
\item for alumina, when one operator and one indentation machine
is used (Fig.~\ref{fi:ident_new1_col_al2o3}-top) and two operators
and two different machines
(Fig.~\ref{fi:ident_new1_col_al2o3}-bottom). A very good agreement
is obtained in both cases and no significant deviation is
observed when two operators and two indentors are used when
compared to a single set of measurements.

\item
for silicon nitride when one operator and one indentation machine
is used (Fig.~\ref{fi:ident_new1_col_si3n4}). Note here that
despite the fact that the crack lengths are characterized by a low
$c/a$ ratio, the experimental results are very well described by
the analytical expression~(\ref{eq:2Dc}) obtained while setting
$m$ to $3/2$ only valid {\it a priori} for $c/a>2$.

\item
for silicon carbide when one operator and one indentation machine
is used (Fig.~\ref{fi:ident_new1_col_sic}-top) and two operators
and two different machines
(Fig.~\ref{fi:ident_new1_col_sic}-bottom). It is noted that
the two sets of results are again very close.

\item
for glass when one operator and one indentation machine is used
(Fig.~\ref{fi:ident_new1_col_glass_all}-top) and two operators and
two different machines
(Fig.~\ref{fi:ident_new1_col_glass_all}-bottom). It is noted
that the two sets of results are also very close.

\end{itemize}

This first identification stage gives very good results. For all
materials, the crack length distributions obtained for various
loads collapse onto a single master curve. Comparing
Figs.~\ref{fi:ident_new1_col_al2o3},
\ref{fi:ident_new1_col_si3n4}, \ref{fi:ident_new1_col_sic},
\ref{fi:ident_new1_col_glass_all} one observes that the width of
the distributions varies noticeably from material to material. The
median crack length $c_m/c^*$ such that $Q(c_m)=0.5$ is
measured to be $c_m/c^* \approx 1.7$ for Al$_2$O$_3$, $1.35$ for
Si$_3$N$_4$, $1.5$ for SiC and $1.15-1.2$ for glass. As discussed
above, the width of the crack length distribution depends upon two
parameters. It first decreases with the system size $(c^*/\xi)$ so
that the finer the microstructure (or the smaller the toughness)
the narrower the distribution. Second, it depends on the width of
the toughness disorder at the microscopic scale. The stronger the
toughness disorder, the larger the distribution. In the present
case, not surprisingly, glass which is characterized by a low
toughness and a very small correlation length for the disorder
gives the narrowest distribution. Conversely the largest
distribution is obtained for the alumina ceramic which has the
coarser microstructure and the highest toughness of the tested
materials (except silicon nitride).

\subsubsection{Step 2}

\begin{figure}[ht!]
%\centerline{\epsfxsize=.6\hsize
%\epsffile{Al2O3/ident_bis_new1_col_al2o3.eps}}
%\centerline{\scalebox{0.45}{\includegraphics{Al2O3/ident_bis_new1_col_al2o3.eps}}}
\centerline{\epsfig{file=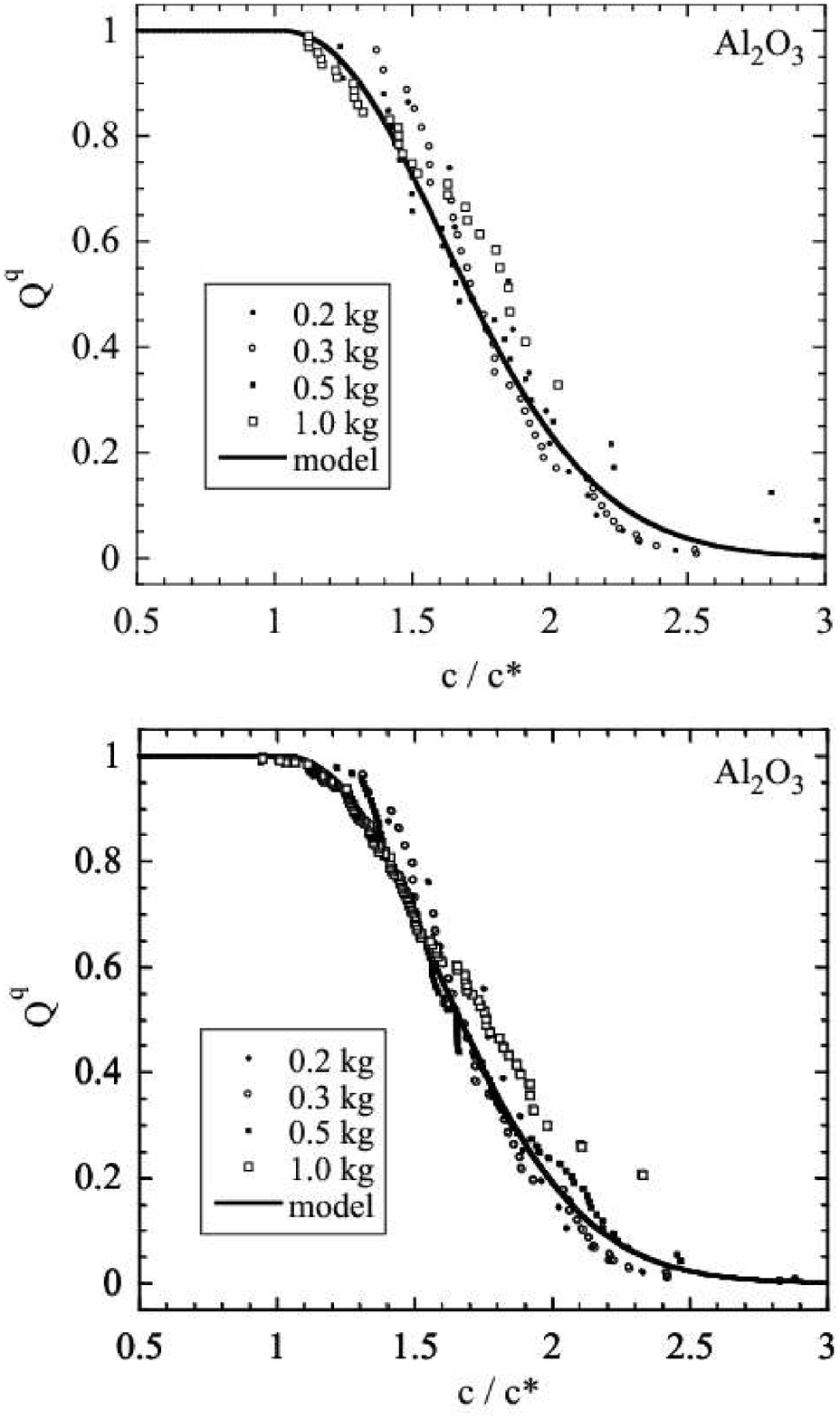,width=0.95\hsize}}

\caption{Rescaled propagation probability $Q^q$ versus
dimensionless crack radius $c/c^*$ for four different applied
masses on alumina. The symbols are experimental data and the line
is the result of the identification with a constant parameter $A$.
>From the present analysis, it is expected that all experimental
points should fall on the same curve. Top: 1 machine and 1
operator. Bottom: 2 indentors and 2 operators.}
\label{fi:ident_bis_new1_col_al2o3}
\end{figure}

\begin{figure}[ht!]
%\centerline{\epsfxsize=.9\hsize
%\epsffile{Si3N4/ident_bis_new1_col_si3n4.eps}}
%\centerline{\scalebox{0.7}{\includegraphics{Si3N4/ident_bis_new1_col_si3n4.eps}}}
 \centerline{\epsfig{file=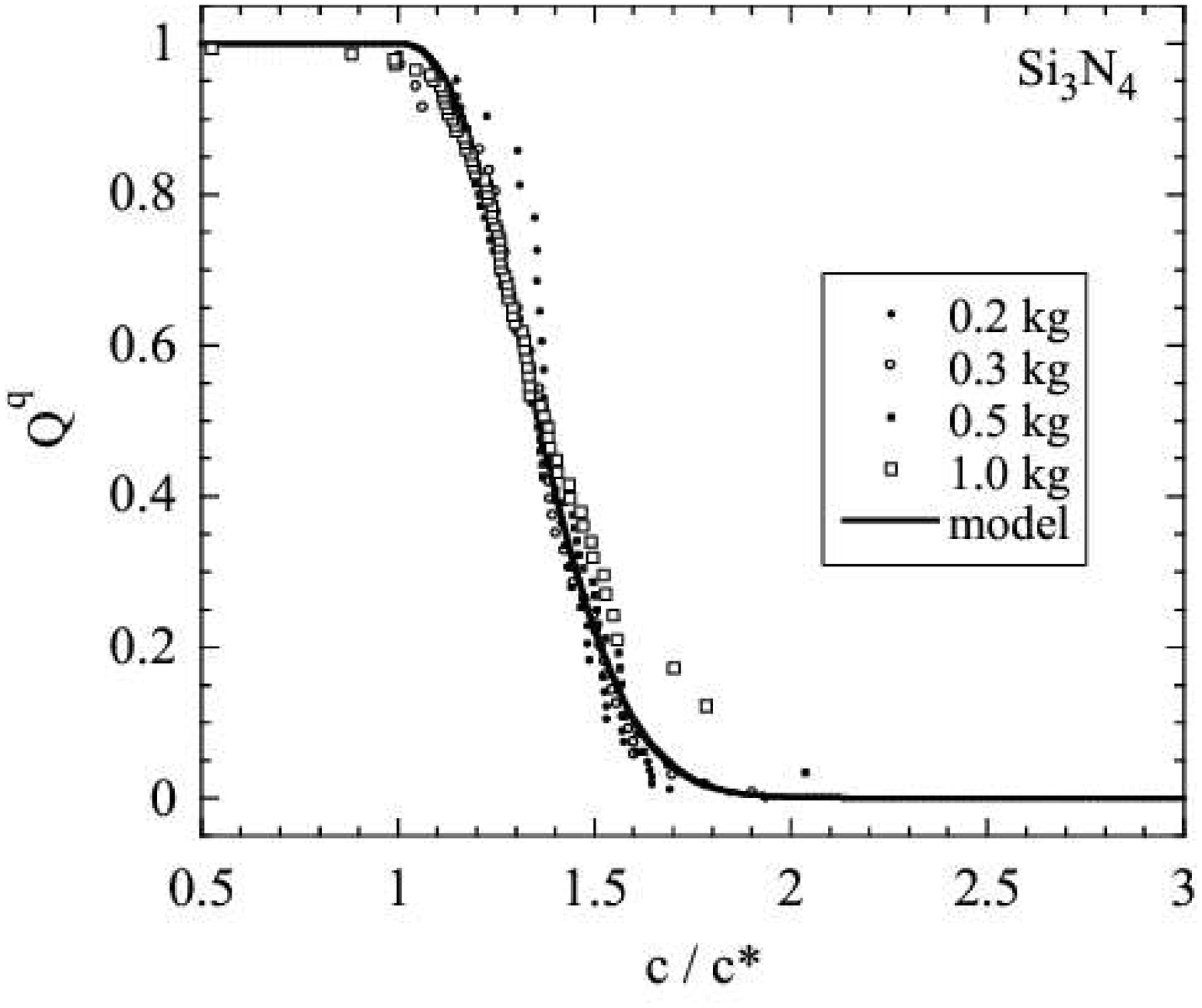,width=0.95\hsize}}

\caption{Rescaled propagation probability $Q^q$ versus
dimensionless crack radius $c/c^*$ for four different applied
masses on silicon nitride. The symbols are experimental data and
the line is the result of the identification with a constant
parameter $A$. From the present analysis, it is expected that all
experimental points should fall on the same curve.}
\label{fi:ident_bis_new1_col_si3n4}
\end{figure}

\begin{figure}[ht!]
%\centerline{\epsfxsize=.6\hsize
%\epsffile{SiC/ident_bis_new1_col_sic.eps}}
%\centerline{\scalebox{0.45}{\includegraphics{SiC/ident_bis_new1_col_sic.eps}}}
  \centerline{\epsfig{file=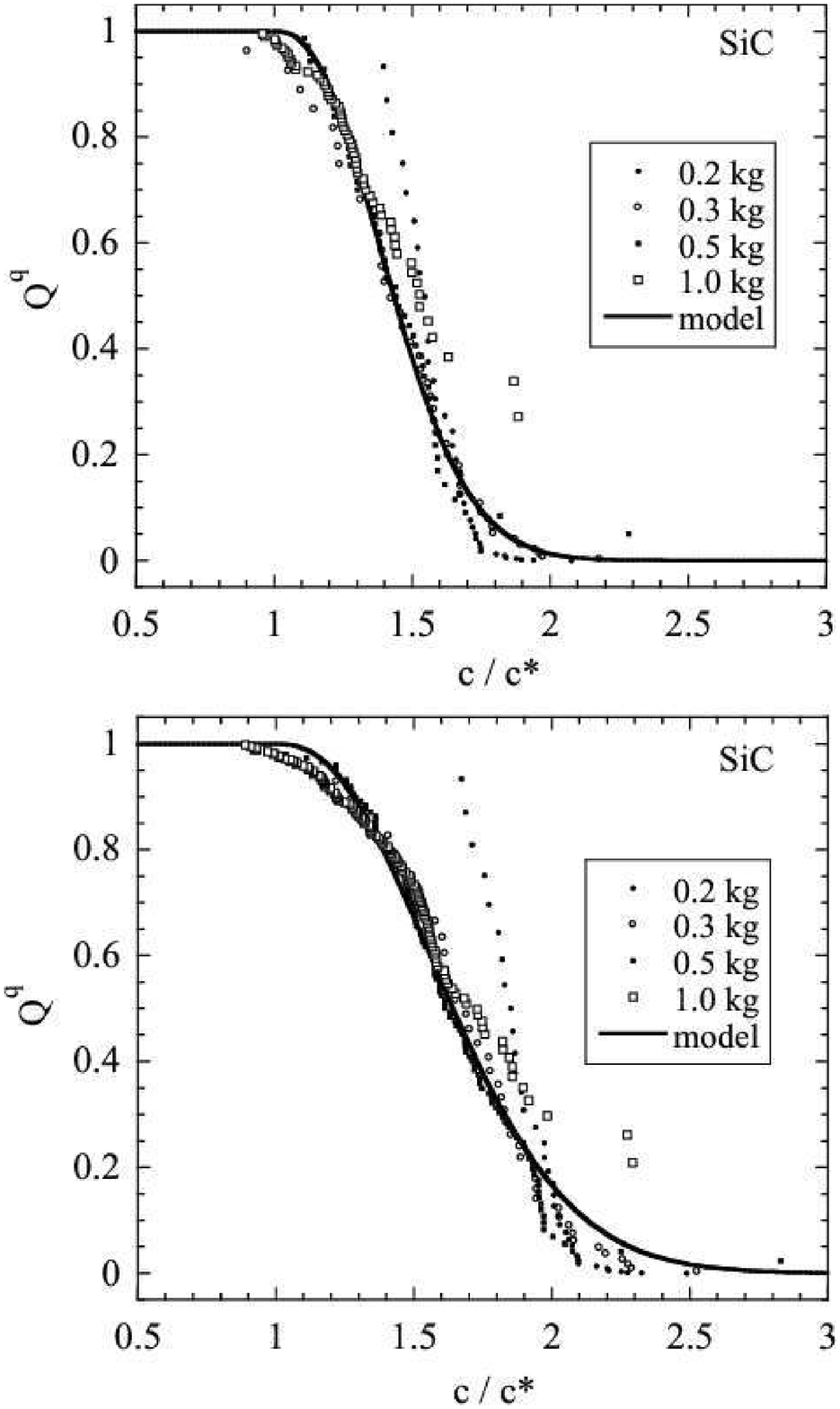,width=0.95\hsize}}

\caption{Rescaled propagation probability $Q^q$ versus
dimensionless crack radius $c/c^*$ for four different applied
masses on silicon carbide. The symbols are experimental data and
the line is the result of the identification with a constant
parameter $A$. From the present analysis, it is expected that all
experimental points should fall on the same curve. Top: 1 machine
and 1 operator. Bottom: 2 indentors and 2 operators.}
\label{fi:ident_bis_new1_col_sic}
\end{figure}

\begin{figure}[ht!]
%\centerline{\epsfxsize=.6\hsize
%\epsffile{verre/ident_bis_new1_col_glass_all.eps}}
%\centerline{\scalebox{0.45}{\includegraphics{verre/ident_bis_new1_col_glass_all.eps}}}
\centerline{\epsfig{file=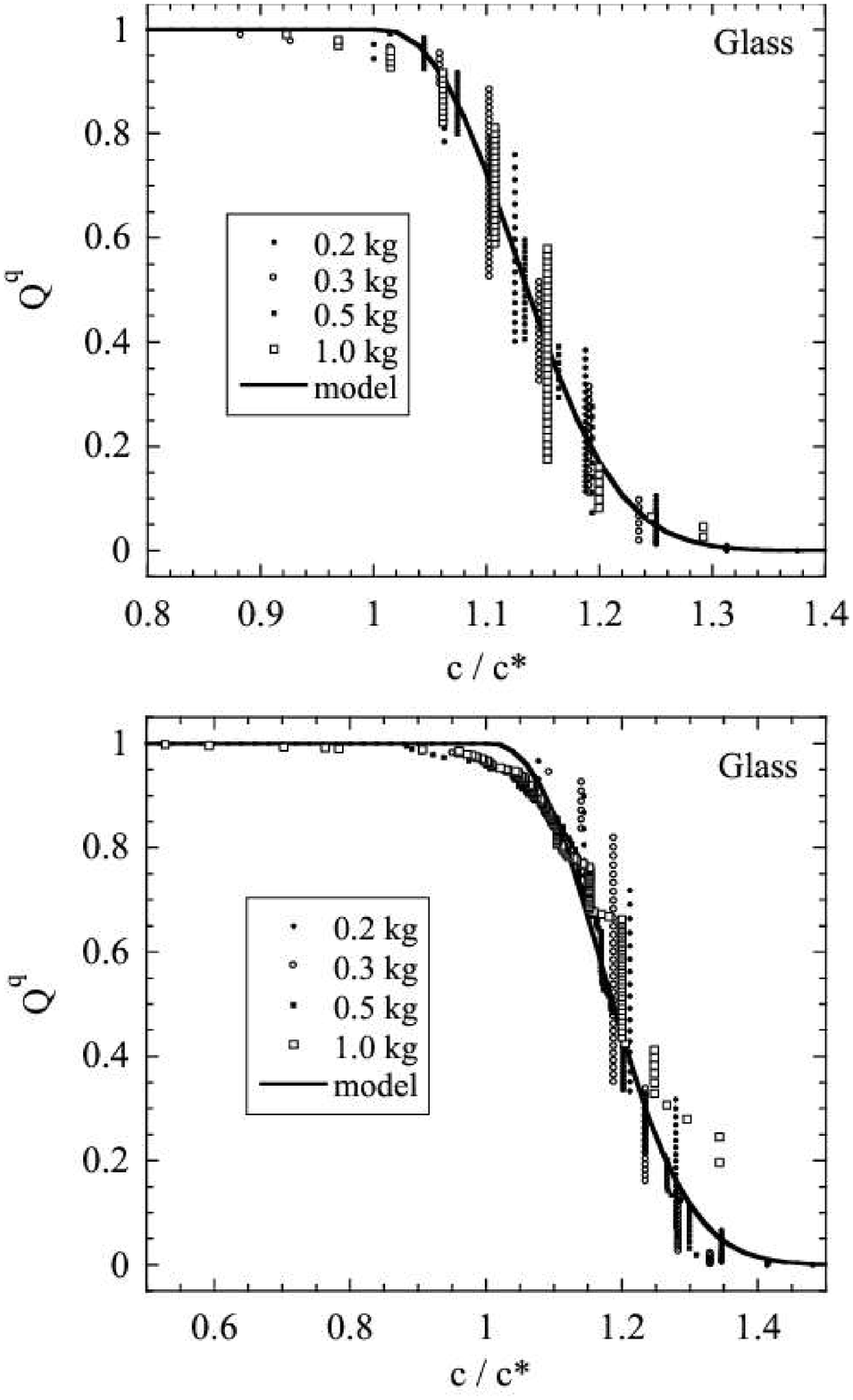,width=0.95\hsize}}
\caption{Rescaled propagation probability $Q^q$ versus
dimensionless crack radius $c/c^*$ for four different applied
masses on sodalime silicate glass. The symbols are experimental
data and the line is the result of the identification with a
constant parameter $A$. From the present analysis, it is expected
that all experimental points should fall on the same curve. Top: 1
machine and 1 operator. Bottom: 2 indentors and 2 operators.}
\label{fi:ident_bis_new1_col_glass_all}
\end{figure}

By using the second identification stage, one obtains the
following results:

\begin{itemize}

\item
for alumina (Fig.~\ref{fi:ident_bis_new1_col_al2o3}) the
uncertainty on $c^*$ is $3~\mu$m when the same machine is used,
and less than $4~\mu$m when the two different machines are
utilized.

\item
for silicon nitride when one operator and one indentation machine
is used (Fig.~\ref{fi:ident_bis_new1_col_si3n4}), an uncertainty
less than $3~\mu$m on $c^*$ is found.

\item
for silicon carbide when one operator and one indentation machine
is used (Fig.~\ref{fi:ident_bis_new1_col_sic}-top), and two
different machines with two operators
(Fig.~\ref{fi:ident_bis_new1_col_sic}-bottom). When the same
parameter $A$ is considered for the three different experimental
conditions, an uncertainty less than $3~\mu$m on $c^*$ is found
when the same machine is used, and less than $5~\mu$m when two
different machines are utilized. The first load level is not well
described. Again this is explained by the fact that, the
ratios $c/a$ being in this case very close to 2, a competition is
observed between the Palmqvist and the radial/median crack
systems. It is thus likely that even if the dependence between
$c^*$ and $M$ is approached by an adapted power law
relationship, the data corresponding to the two crack systems
cannot be treated together.

\item
for glass when one operator and one indentation machine is used
(Fig.~\ref{fi:ident_bis_new1_col_glass_all}-top), and two
operators and two different machines
(Fig.~\ref{fi:ident_bis_new1_col_glass_all}-bottom). When the same
parameter $A$ is considered for the two different experimental
conditions, an uncertainty less than $7~\mu$m on $c^*$ is found.

\end{itemize}

\subsubsection{Step 3}

\begin{figure}[ht!]
%\centerline{\epsfxsize=.9\hsize
%\epsffile{cstar_ident_bis_new1.eps}}
%\centerline{\scalebox{0.7}{\includegraphics{cstar_ident_bis_new1.eps}}}
\centerline{\epsfig{file=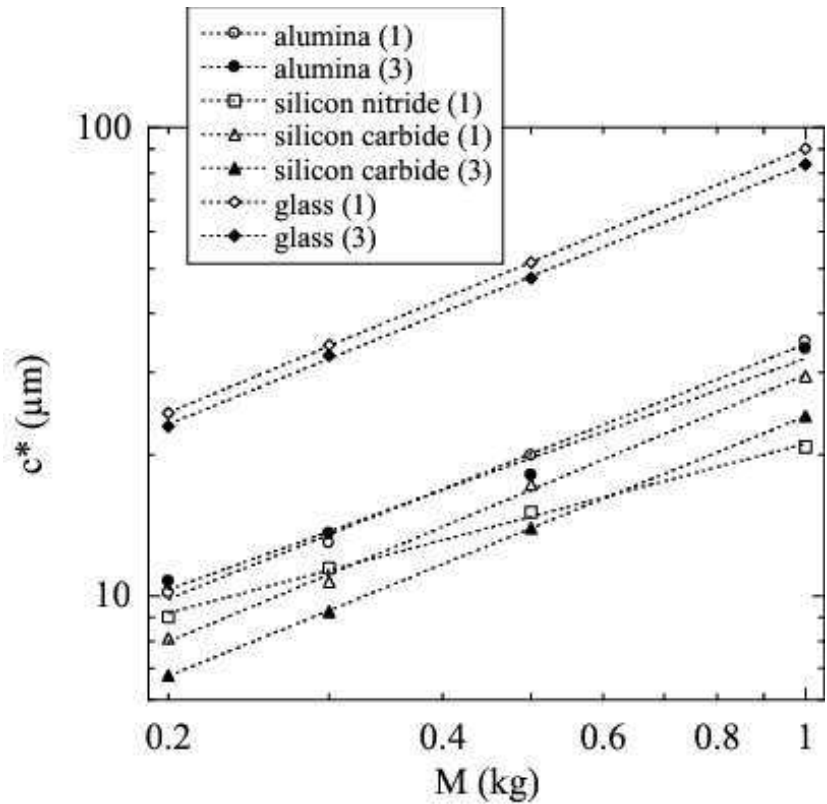,width=0.95\hsize}}

\caption{Parameter $c^*$ versus applied mass $M$ for the six
series of experiments. A power law fits reasonably all the
experiments (dashed lines). The number in parentheses indicates
the number of situations (machine and operator) used to perform
the identification} \label{fi:cstar_ident_bis_new1}
\end{figure}

The self-consistency was checked for all the situations.
Figure~\ref{fi:cstar_ident_bis_new1} shows the dependence of the
characteristic radius $c^*$ with the applied mass $M$. When a
power law dependence is sought, all conditions apart from silicon
nitride have approximately the same exponent (\ie $1/m'\approx
0.8$ compared to the expected asymptotic value 0.67). When the
fixed-point algorithm is used, the difference between the first
estimate of $m'$ and its converged value is always less than 4\%
for all the analyzed cases. Furthermore, convergence is very
fast since at most three iterations were used to get the final
results.

\section{Discussion}

The present analysis was obtained using two main hypotheses, namely,
an elastic-brittle behavior and strong pinning conditions.
Moreover the crack system was assumed to be radial/median. In the
following, the results of the statistical analysis are discussed
in the light of these different hypotheses.

\subsection{nature of the crack system}
\label{palmqvist}

\begin{figure}[ht!]
\centerline{\epsfig{file=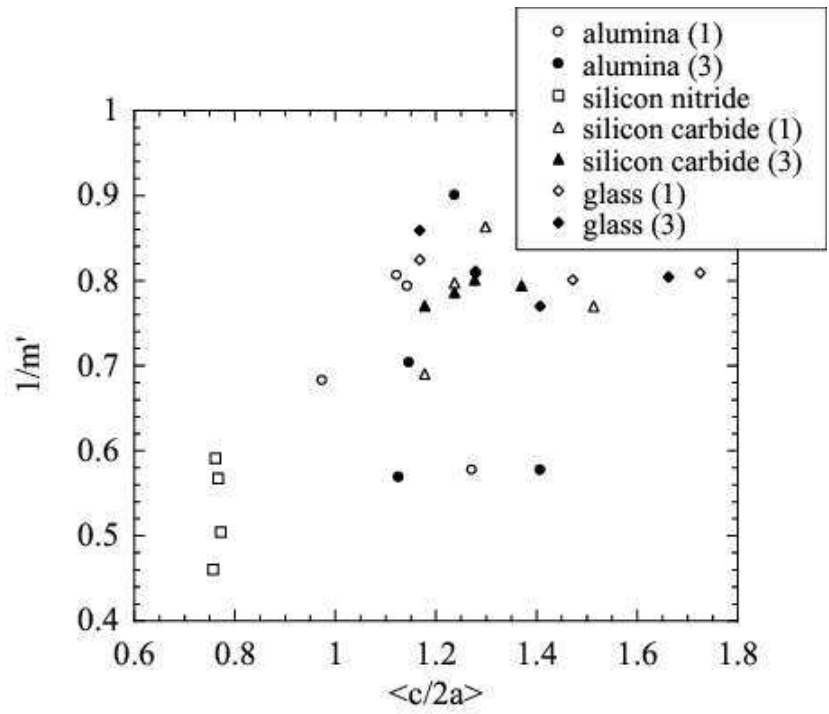,width=0.95\hsize}}

\caption{Change of $1/m'$ versus $<c/2a>$ for all the
experiments.} \label{fi:m_vs_c_2a}
\end{figure}

Table~1 summarizes the results obtained for the effective exponent
$m$ and the ratio $c/a$ for the various materials and experimental
conditions. The same results are presented in
Fig.~\ref{fi:m_vs_c_2a}. Looking at the dependence of the measured
exponent $1/m'$ on the average $c/2a$ value, one clearly
distinguishes two groups of materials. The first one consists of
silicon nitride and the other one of glass and silicon carbide. As
described above average crack length values for silicon nitride
are less than the other ones. Moreover exponent values are also
lower and close to 0.5-0.6. Conversely, SiC~(1), SiC~(3),
glass~(1) and glass~(3) exponents all concentrate close to 0.8,
and seem to approach closer to this value when the ratio $\langle
c/2a\rangle$ increases (even if this change is not monotonic). The
case of alumina is more dubious, even the results corresponding to
large $c/a$ values do not exhibit a clear belonging to any of
these two groups.

These separate behaviors are attributed to the nature of the crack
system. As discussed above, the present study is developed in the
framework of a median/radial crack system. However, the Palmqvist
crack system classically appears for high toughness materials. In
the case of low toughness materials, the crack system is
predominantly median/radial for high loads but may be of Palmqvist
type if the applied load is low enough. Experimentally, it is
difficult to discriminate between these two crack systems without
direct observation. In particular, the value of the ratio $c/2a$
is often not sufficient to predict the generated
indentation-crack~\citep{Glandus-Rouxel-GLA91}. Beyond the
influence on the ratio $c/a$, the nature of the crack system also
affects the scaling of the crack length with load, namely, a
Palmqvist crack system generates the scaling $l \propto M$ (with
$l=c-a$), while a radial one induces $c \propto M^{2/3}$. For
silicon nitride, $c/2a$ values obtained during indentation
experiments, and the unusual values for $1/m'$, allow us to
conclude that the generated crack system is of Palmqvist type,
namely, for such a material, \citet{Wang02} have reported a change
in the crack system for $c/a$ between 2.3 and 2.4.

\begin{figure}[ht!]
\centerline{\epsfig{file=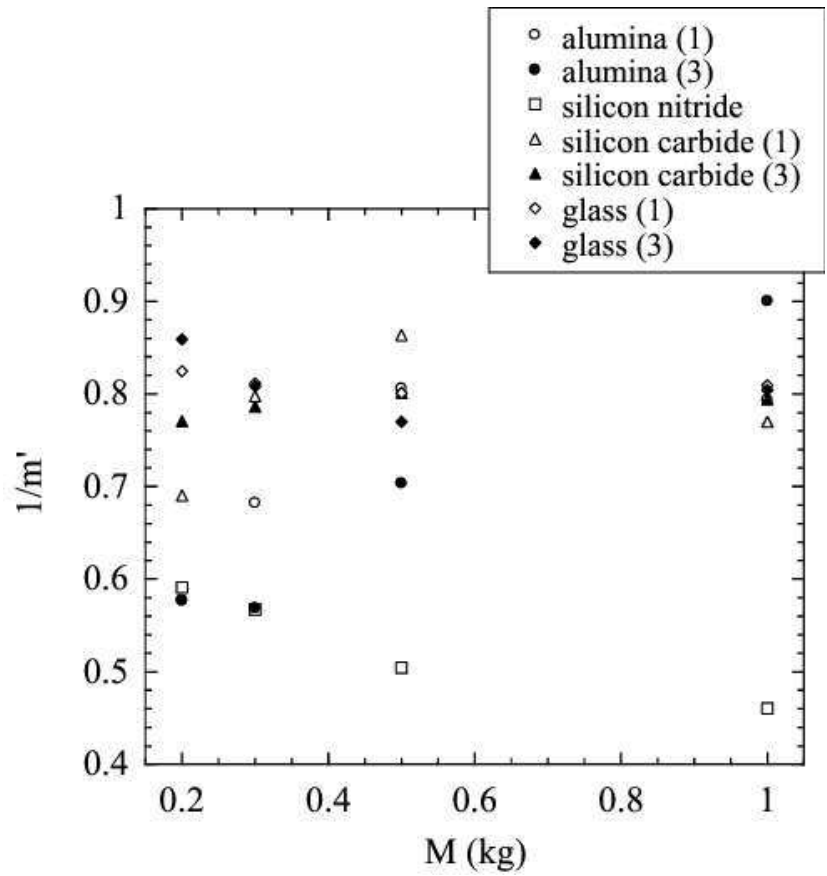,width=0.95\hsize}}

\caption{Change of $1/m'$ versus applied mass $M$ for all the
experiments.} \label{fi:m_vs_F}
\end{figure}

The dependence of the scaling behavior on the nature of the crack
system is seen again when plotting the exponent $m$ against the
applied mass (Fig.~\ref{fi:m_vs_F}). One observes first that the
two above identified groups still exist, namely, silicon nitride
exponents exhibit a (decreasing) convergence toward 0.4-0.45 when
the applied mass increases, while both glass and silicon carbide
exponents converge toward 0.8. The previous conclusions on the
generated crack system for these materials are therefore
confirmed.

Let us now discuss the behavior of alumina exponents. Namely,
when the applied mass increases, so does $1/m'$ from silicon
nitride exponent values to glass and silicon carbide ones. One
concludes that for high applied loads, the indentation generated
crack system is the same as for glass (or silicon carbide). Yet,
for low applied loads, this system is closer to a Palmqvist or a
mixed system. To conclude on the latter crack systems as a
function of the applied mass, one would have to perform
additional experiments with both lower and higher load levels,
and direct observation of the developed cracks should be carried
out.

By considering the results obtained for silicon carbide, glass,
and partially for alumina, one concludes that the identified
values for the exponent between $c^*$ and the applied mass $M$
are consistent with both theory and observations, even if one has
to account for an ``asymptotic'' $1/m'$ value for radial crack
systems greater than the expected one (0.8 instead of 0.67). The
universal distribution proposed to describe the statistics of
indentation crack lengths gives a satisfactory account of most of
the experimental data obtained on different materials. Moreover
the few experimental data not properly described by this approach
developed in the framework of a median/radial crack system may
be attributed to Palmqvist indentation cracks.\\

\subsection{Microscopic interpretation}

As discussed above, the statistical analysis allows us to extract
a microscopic parameter $A$ that is re-expressed as a typical
length, namely, $A^{1/\zeta-2}=A_0^{1/\zeta-2}\xi
\sigma_0^{1/1-\zeta}$. This expression includes the
characteristic size $\xi$ of the heterogeneities as well as a
factor dependent on the level of the local disorder. Table~2
summarizes the results obtained for the different materials
studied herein. Let us compare them with estimated characteristic
scales of the structural disorder. Apart from the case of glass,
these results are consistent with relative toughness fluctuations
of order 1. Note however that these results have to be considered
cautiously. Two hypotheses of the analysis may affect the
interpretation of the parameter $A$ and more specifically the
length scale $\xi$. The assumption of an elastic-brittle behavior
imposes one to consider crack fronts of extension greater than the
size of the process zone. The latter is thus a lower bound for
$\xi$ even if a structural disorder may exist at finer scales. A
more questionable point is the hypothesis of strong pinning. The
latter corresponds to situations where the structural disorder is
strong enough to locally arrest the crack front. As discussed
above, this induces an intermittent dynamics of the crack
propagation. Conversely, in weak pinning conditions, the toughness
disorder only modulates the front conformation and crack fronts
change smoothly. It is shown that strong pinning conditions are
always obtained in the limit of large fracture
fronts~\citep{RVH-EJMA03}. In the case of a rather weak structural
disorder, one thus may expect that the characteristic length $\xi$
no longer be defined by the characteristic size of the
heterogeneities but by a larger scale corresponding to the
transition between weak and strong pinning.

\section{Summary}

By using a model developed to characterize the statistical
properties of a crack front propagating through a heterogeneous
material, an analytical expression is given for the distribution
of crack arrest lengths. The latter was shown to be
material-independent. The distribution presents a universal shape and is
fully characterized by two parameters. The first one corresponds
to the macroscopic toughness value. The second one, which gives
the width of the distribution, depends on the relative size of the
cracks compared the size of the microstructure and on the width of
the microscopic toughness disorder (for very large sizes, the
distribution tends to a Dirac centered on the macroscopic
toughness value).

Experimental tests have been performed on four different materials
(alumina, silicon nitride, silicon carbide and sodalime silicate
glass) with four different loads (0.2~kg, 0.3~kg, 0.5~kg and
1~kg). A particular attention was given to potential sources of
non-intrinsic fluctuations, namely the dependence on the operator
and on the testing machine. It appeared that the analytical
expression gives a very good account of the data. Coupled with
experimental observations the statistical analysis allowed us to
discriminate between indentation cracks belonging to a
median/radial and a Palmqvist crack system, respectively.
Restricting ourselves to the former case within which the present
analysis was developed, fitting parameters consistent with the
scaling laws expected in the geometry of indentation tests were
obtained.

A very strong feature of the results obtained in the present work
is that the knowledge of the crack length (or effective toughness)
distribution at a given scale gives an immediate access to the
distributions corresponding to any other scale. In particular,
this should help improving the quality of crack length or
toughness estimates for large systems when using data obtained at
microscopic or a much finer scale.

%################################################################
%################################################################
%
\section*{Acknowledgements}
The authors acknowledge useful discussions with Pr.~Tanguy Rouxel and
Dr.~Ren\'e~Gy.
%
%################################################################
%################################################################
%

\newpage

\begin{table}[ht!]
\label{table-m-vs-ca}
\begin{center}
    \caption{Identified exponent of the power law between $c^*$ and the applied mass $M$ and average
    normalized crack length for all studied materials and masses (expressed in kg),
    for (1) one operator and one indentation machine,
    and (3) two operators and two indentation machines}
     \label{tab:res_power_law_ident}
     \begin{tabular}{|c|c||c|c|c|c|c|c|c|}
        \hline
    Mass &    \multirow{2}{1.5cm} {Parameters} & \multicolumn{2}{|c|}{Al$_2$0$_3$} & \multirow{2}{1cm} {Si$_3$N$_4$} & \multicolumn{2}{|c|}{SiC}  & \multicolumn{2}{|c|}{Glass} \\
       \cline{3-4} \cline{6-9}
      (kg) & & (1) & (3) &  &  (1) & (3) & (1) & (3)\\
      \hline \hline
           \multirow{2}{.5 cm} {1.0}  & $1/m'$ & 0.79 & 0.90 & 0.46 & 0.77 & 0.79 & 0.81 & 0.80\\
            \cline{2-9}
            & $<c/2a>$ & 1.41 & 1.24 & 0.76 & 1.51 & 1.37 & 1.72 & 1.66\\
            \hline
          \multirow{2}{.5 cm} {0.5}   &$1/m'$  & 0.84 & 0.70 & 0.50 & 0.86 & 0.80 & 0.80 & 0.77\\
            \cline{2-9}
           & $<c/2a>$ & 1.13 & 1.15 & 0.77 & 1.30 & 1.28 & 1.47 & 1.41\\
            \hline
          \multirow{2}{.5 cm} {0.3}   &$1/m'$  & 0.68 & 0.57 & 0.57 & 0.80 & 0.79 & 0.81 & 0.81\\
            \cline{2-9}
           &  $<c/2a>$ & 1.27 & 1.13 & 0.77 & 1.24 & 1.24 & 1.28 & 1.28\\
            \hline
          \multirow{2}{.5 cm} {0.2}  &$1/m'$ & 0.58 & 0.58 & 0.59 & 0.69 & 0.77 & 0.82 & 0.86\\
            \cline{2-9}
            & $<c/2a>$ & 1.31 & 1.41 & 0.76 & 1.178 & 1.178 & 1.17 & 1.17\\
            \hline
     \end{tabular}
\end{center}
\end{table}

\newpage

\begin{table}[ht!]
  \begin{center}
    \caption{Identified parameters $A$ and estimation of the characteristic
scales of the structural disorder for the four tested materials.}
     \begin{tabular}{|c|c||c|c|c|c|c|c|c|}
        \hline
    Mass &   Parameters & \multicolumn{2}{|c|}{Al$_2$0$_3$} & \multirow{2}{1cm} {Si$_3$N$_4$} & \multicolumn{2}{|c|}{SiC}  & \multicolumn{2}{|c|}{Glass} \\
       \cline{3-4} \cline{6-9}
      (kg) &  & (1) & (3) &  &  (1) & (3) & (1) & (3)\\
      \hline \hline
           - & $\xi$ ($\mu$m) & \multicolumn{2}{|c|}{10} & 3 &  \multicolumn{2}{|c|}{4} &  \multicolumn{2}{|c|}{0.002} \\
           \hline
           1.0  & \multirow{4}{0.5cm} {$A^{\frac{1} {2-\zeta}}$ ($\mu$m$^{-1}$)} & 0.08& 0.12& 0.34& 0.17& 0.14& 0.32& 0.18\\
            \cline{1-1} \cline{3-9}
           0.5  & & 0.12& 0.15& 0.39& 0.29& 0.25& 0.65& 0.42\\
            \cline{1-1} \cline{3-9}
           0.3  & & 0.19& 0.19& 0.50& 0.29& 0.29& 0.75& 0.75\\
            \cline{1-1} \cline{3-9}
           0.2  &  & 0.22& 0.22& 0.80& 0.88& 0.88& 0.62& 0.62\\
            \hline
     \end{tabular}
  \end{center}
\label{micro}
\end{table}

\newpage
%\vfill
%%%%%%%%%%% 1
%

\newpage
%%%%%%%%%%% 2
%
%\null \vfill1cm
\vfill
\newpage
%
%%%%%%%%%%%% 3
%

\newpage
%
%
%%%%%%%%%%%% 4
%

\vfill
%\newpage
%%%%%%%%%%%% 5
%

\newpage
%%%%%%%%%%%% 6
%
%

\vfill
%\newpage
%
%
%%%%%%%%%%%% 7
%
\vfill

\newpage
%
%%%%%%%%%%% 8
%
\vfill
\newpage
%
%%%%%%%%%%%% 9
%
\vfill
\newpage
%
%%%%%%%%%%%% 10
%
\vfill
\newpage
%
%
%%%%%%%%%%%% 11
%
%
\newpage

%%%%%%%%%%%% 12
%
%
%
\newpage
%
%%%%%%%%%%%% 13
%
%
\newpage
%
%
%%%%%%%%%%%% 14
%
%
%
\newpage
%
%%%%%%%%%% 15
%

\newpage
%
%%%%%%%%% 16
%
\newpage
%
%%%%%%%%%%% 17
%
\vfill
%\newpage
%
%%%%%%%%%%%% 18
%

\end{document}